\documentclass{article} % For LaTeX2e
\usepackage{iclr2025_conference,times} %iclr2025_conference,
\usepackage{amsmath,amsfonts,bm}

\def\eqref#1{equation~\ref{#1}}
\def\1{\bm{1}}

\DeclareMathAlphabet{\mathsfit}{\encodingdefault}{\sfdefault}{m}{sl}
\SetMathAlphabet{\mathsfit}{bold}{\encodingdefault}{\sfdefault}{bx}{n}

\usepackage{graphicx}%
\usepackage{hyperref}
\usepackage{url}
\usepackage[utf8]{inputenc} % allow utf-8 input
\usepackage[T1]{fontenc}    % use 8-bit T1 fonts           % simple URL typesetting
\usepackage{booktabs}
\usepackage{tabularx}% professional-quality tables
\usepackage{amsfonts}       % blackboard math symbols
\usepackage{nicefrac}       % compact symbols for 1/2, etc.
\usepackage{xcolor}         % colors
\usepackage{graphicx}% professional-quality tables
\usepackage{microtype}      % microtypography
\usepackage{xcolor}         % colors
\usepackage{amsmath}
\usepackage{enumitem}
\usepackage{wrapfig}
\usepackage{adjustbox}
\title{Brain-to-Text Decoding with Context-Aware Neural Representations and Large Language Models}
\author{
  Jingyuan Li\thanks{Authors contributed equally to this work.}\\
  Department of Electrical and Computer Engineering\\
   University of Washington \\
  \texttt{jingyli6@uw.edu} \\
 \And
  Trung Le$^*$\\
  Department of Electrical and Computer Engineering\\
  University of Washington \\
  \texttt{tle45@uw.edu} \\
  \And
   Chaofei Fan\\
   Department of Computer Science\\
  Stanford University\\
  \texttt{stfan@stanford.edu} \\
  \And
     Mingfei Chen\\
   Department of Electrical and Computer Engineering\\
  University of Washington\\
  \texttt{lasiafly@uw.edu } \\
  \And
   Eli Shlizerman\\
  Department of Applied Mathematics\\ Department of Electrical and Computer Engineering\\
  University of Washington\\  \texttt{shlizee@uw.edu} \\
}
\iclrfinalcopy % Uncomment for camera-ready version, but NOT for submission.
\begin{document}

\maketitle
\begin{abstract}
Decoding attempted speech from neural activity offers a promising avenue for restoring communication abilities in individuals with speech impairments. Previous studies have focused on mapping neural activity to text using phonemes as the intermediate target. 
While successful, decoding neural activity directly to phonemes ignores the context dependent nature of the neural activity-to-phoneme mapping in the brain, leading to suboptimal decoding performance.
% Such strategy was shown to lead to more optimal decoding than direct neural activity to text mapping, however benchmarks indicate that there is room for improvement in decoding accuracy. One of the reasons for this sub-optimality is that phonemes do not capture the full context of speech in neural activity. 
%To address this challenge, 
In this work, we propose the use of diphone - an acoustic representation that captures the transitions between two phonemes - as the context-aware modeling target. We integrate diphones into existing phoneme decoding frameworks through a novel divide-and-conquer strategy in which we model the phoneme distribution by marginalizing over the diphone distribution. Our approach effectively leverages the enhanced context-aware representation of diphones while preserving the manageable class size of phonemes, a key factor in simplifying the subsequent phoneme-to-text conversion task. We demonstrate the effectiveness of our approach on the Brain-to-Text 2024 benchmark, where it achieves state-of-the-art Phoneme Error Rate (PER) of 15.34\% compared to 16.62\% PER of monophone-based decoding. When coupled with finetuned Large Language Models (LLMs), our method yields a Word Error Rate (WER) of 5.77\%, significantly outperforming the 8.93\% WER of the leading method in the benchmark.
\end{abstract}

\section{Introduction}

Verbal communication is a unique feature of human social interaction. Loss of ability to articulate speech as a result of neurological pathologies such as stroke and Amyotrophic Lateral Sclerosis (ALS) can significantly reduce the quality of life for affected individuals. Recent advancements in Brain-Computer Interfaces (BCI) offer promising pathways toward restoring communication ability in these patients by translating neural activity into communicative messages. These messages can be conveyed through various modalities, including typed characters \citep{pandarinath2017high}, handwriting \citep{willett2021high}, text \citep{herff2015brain, willett2023high, metzger2023high}, and synthesized speech \citep{metzger2023high}. 

Among existing speech BCI systems, the methods with highest decoding accuracy and throughput are those that translate neural signals associated with orofacial movements during attempted speech into fundamental acoustic units (phonemes), which are then decoded into words and sentences~\citep{willett2023high, metzger2023high, card2024accurate}. This two-staged approach typically involves (1) neural signal to phonemes: using a temporal deep network to decode a binned multi-channel neural time series into probability of phonemes being spoken at each time step, and (2) phonemes to text: employing a language model (LM) to infer the most probable sequence of words given the phoneme probabilities. 

Prior work shows that decoding phonemes as an intermediate representation rather than directly decoding words, provides the system the flexibility to decode phrases from extensive vocabularies a limited set of training examples~\citep{metzger2023high}, since from a fixed set of 40 phonemes, one can practically construct any word of any arbitrary length. This scalability is especially advantageous given the limited availability of neural recordings in clinical settings. Indeed, prior works and state of the art methods show that such an intermediate step is advantageous. 
\begin{figure*}
%\begin{center}
\begin{center}
 \includegraphics[width=0.9\linewidth]{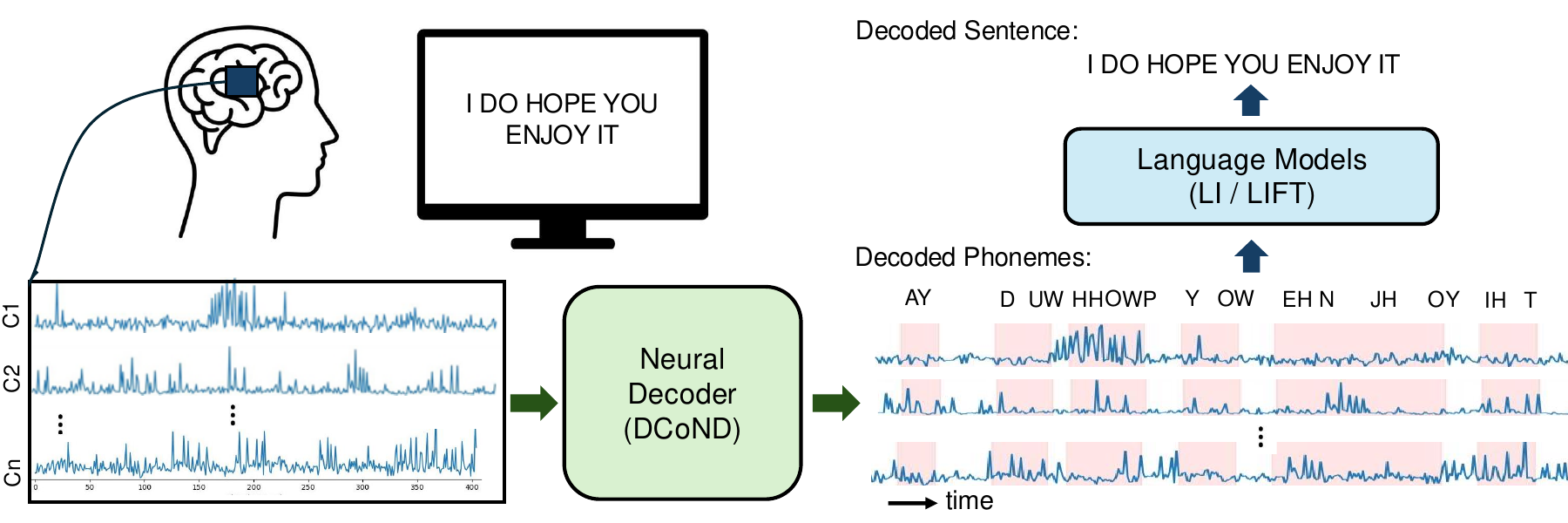}
\end{center}
   \caption{Overview of the Brain-to-Text decoding pipeline. The Neural Decoder with Divide-and-Conquer Strategy (DCoND) decodes multi-channel neural activity into phonemes. The phonemes are subsequently converted into words by LLMs using either ICL or fine-tuning techniques.}
\label{fig:fig1-task-illustraion}
\vspace{-10pt}
\end{figure*}

While decoding single phonemes from neural activity may offer more scalability than direct decoding of words, it remains a challenging task. Given the innate variability of neural signals, the mapping from neural activity to phonemes is many-to-one and highly nonlinear. Furthermore, evidence suggests that cortical activation patterns producing a particular phoneme are not static, but can vary depending on the context of surrounding phonemes, a phenomenon known as \textit{coarticulation}~\citep{bouchard2014neural, mugler2014direct}. In other words, cortical neurons at any given time during speech production are likely encoding a phoneme along with its context, rather than a phoneme in isolation. Given this observation, diphone \citep{nedel2000phone} - a sequence of two adjacent phonemes - is a more suitable representation for capturing this context dependency in neural signals and potentially reducing the nonlinearity in phoneme decoding. Hence we propose to decompose the phoneme classification task into the subtasks of 1) diphone classification, after which 2) diphone probabilities are summed up to obtain the phoneme prediction, i.e. predicting phoneme distribution by marginalizing over the diphone distribution. We show that this divide-and-conquer strategy significantly enhances phoneme decoding performance. 

Recently introduced approaches leverage language models, such as n-gram model, to translate phoneme probabilities into words \citep{willett2023high, metzger2023high, benster2024cross}. Notably, \citet{benster2024cross} further uses GPT3.5~\citep{brown2020language} after a 5-gram model to refine the resulting word sequences into coherent sentences by ensembling multiple 5-gram transcription candidates. However, the transcription candidates generated by the n-gram model can significantly deviate from the ground truth phoneme sequence. To address this issue, we propose to augment the ensembling method in \citep{benster2024cross} to include decoded phonemes alongside transcription candidates, which proves to provide extra information for GPT3.5 to infer the correct transcription. %Additionally, 
In addition, we propose an In-Context Learning (ICL) paradigm for LLMs, enabling them to adapt quickly to newly decoded inputs in a gradient-free manner without the need for the computationally expensive finetuning process. This approach offers a more efficient alternative for improving transcription accuracy in resource-constrained settings.

In summary, our contributions in this work are as follows:
\begin{itemize}[leftmargin=*]
    \item We propose DCoND (\textbf{D}ivide-and-\textbf{Co}nquer \textbf{N}eural \textbf{D}ecoder), a novel framework for decoding phonemes from neural activity during attempted speech. Backed by neuroscientific insights, DCoND infers the temporal phoneme distribution by marginalizing over the diphone distribution, leveraging the context-dependent nature of phonemes in neural representation. The divide-and-conquer strategy can serve as an adaptable component for various neural decoders in brain-to-text decoding.
    \item We propose incorporating decoded phonemes alongside decoded words in an \textbf{L}LM-based ensembling strategy to enhance the speech decoding performance. We also propose the use of (\textbf{I}CL) paradigm (DCoND-LI) as an alternative to \textbf{F}ine\textbf{T}uning LLMs (DCoND-LIFT), offering a more efficient solution for resource-constrained brain-to-text systems. 
    % \item We propose enhancements to the existing phonemes-to-text pipeline to refine the decoded phonemes and sentences. Our contribution in this part includes three major components: \jingyuan{should we include these detailed explanation?}
    % \begin{itemize}
    %     \item We propose a cascaded LLM pipeline that leverages different aspects of different language models (5-gram, OPT and GPT3.5) for refining transcription predictions.
    %     \item We propose to incorporate decoded phonemes alongside decoded words to enhance the efficiency of GPT3.5 in learning to correct the transcription predictions.
    %     \item Apart from finetuning GPT3.5 as in \citep{benster2024cross}, we additionally propose the use of in-context learning inference in resource-constrained brain-to-text systems, enabling quick adaptation of GPT3.5 based on just a few context examples.
    % \end{itemize}
    \item We demonstrate the effectiveness of our approaches on the Brain-to-Text 2024 benchmark, where our approach outperforms existing state-of-the-art (SOTA) approaches and achieves PER of 15.34\% and WER of 5.77\%, a significant improvement compared to 8.93\% WER of the leading existing SOTA method. 
\end{itemize}

%\annotate{some potential content for intro: Many studies have been focused on deciphering the brain signal into different modalities, including decoding neural signals into stereotypical behavior \citep{azabou2024unified, liu2022seeing, sussillo2016lfads, milstein2017multiscale},  translating neural signals into visual inputs \citep{benchetrit2023brain, lan2023seeing, xia2024dream, sun2024neurocine, defossez2023decoding}, generating the perceived speech signal from neural signal \citep{defossez2023decoding, fodor2024towards, yang2024decode}, or identifying attempted complex behavior such as speech production \citep{willett2023high, metzger2023high}. single phoneme could be easier to classify with higher accuracy compared to the situation when phonemes exist in a words \citep{mugler2014direct}}

\section{Related Work}
\label{related_work}
\paragraph{Brain-to-Text Decoding with Speech Waveforms}
When the temporal correspondence between the neural signal and the speech is known, the problem of decoding speech from neural activity is relatively more manageable. Such a situation occurs during speech perception tasks \citep{poeppel2008speech, defossez2023decoding, fodor2024towards, yang2024decode}. In this case, the mapping from neural activity to perceived speech could be learned through supervised learning \citep{fodor2024towards,yang2024decode} or contrastive learning \citep{defossez2023decoding}.
Temporal correspondence between neural activity and speech also exists in speech production experiments performed by individuals who still retain the ability to speak normally, during which concurrent speech waveforms are recorded. Studies for such scenarios include \citep{jou2006towards, schultz2010modeling, kapur2018alterego, meltzner2018development, diener2018session, janke2017emg, chen2024neural}. When the produced speech is not fully observed, Gaddy and Klein propose to use dynamic time warping and canonical correlation analysis to align the neural signals with recorded audio signal \citep{gaddy2020digital,gaddy2021improved}. In contrast to these works, our study focuses on speech decoding when audio recordings of speech are not available.
% These methods still need example audio signals during training as anchors \citep{benster2024cross}, which are not practical for patients with neurological impairment. While, our study focuses more on situations where audio signals are not available.
\paragraph{Brain-to-Text Decoding without Speech Waveforms}
In cases of individuals who cannot produce intelligible speech, the speech decoding problem could be entirely avoided by using typing-based systems, albeit with low throughput \citep{vansteensel2016fully, pandarinath2017high, linse2018communication}. Early works on speech decoding were demonstrated with a small vocabulary size \citep{moses2021neuroprosthesis,kellis2010decoding}, which could be improved by learning to decode letters \citep{metzger2022generalizable}. Other studies investigate phonemes as the decoding target \citep{ pei2011decoding, mugler2014direct, herff2015brain,willett2023high, metzger2023high}. However, decoding phonemes directly can be a difficult task since neural representations for phonemes could change depending on the contexts they are spoken \citep{mugler2014direct}. We leverage this observation to devise our strategy using diphones as the decoding target. 

\paragraph{Brain-to-Text Decoding vs. Speech-to-Text Decoding}
While brain-to-text and speech-to-text decoding share certain similarities decoding text from neural signals is a significantly more challenging task. A key difference is that speech signals are univariate, while neural activity is multivariate as it is recorded by multi-channel electrodes. Furthermore, neural signals are far more intricate. Less is known about how neurons encode speech within their spiking activity, as well as the degree to which speech-relevant components can be extracted from the complex interaction of neural population. However, brain-to-text decoding methods have drawn inspiration from speech-to-text decoding research, commonly referred to as Automatic Speech Recognition (ASR). Earlier studies \citep{miao2015eesen, aggarwal2011acoustic, huang2014historical} use Hidden Markov Models and Gaussian Mixture Models to decode recorded speech signals into phonemes before translating into words. \citep{darjaa2011rule,laleye2016first} suggest that using diphone or triphone could enhance the accuracy of ASR systems. Modern ASR systems have transitioned to end-to-end learning approaches, directly decoding speech signals into words \citep{prabhavalkar2023end, graves2012sequence, gulati2020conformer, hsu2021hubert, schneider2019wav2vec}. 
%,senior2015acoustic,chorowski2019towards
These systems typically utilize Transformers \citep{dong2018speech, zhang2020transformer} or hybrid architectures that combine Transformers with Convolutional Neural Networks \citep{gulati2020conformer}. 
A common attribute to end-to-end learning with Transformers based architectures is that this type of learning requires large, multiple and diverse input-to-text datasets. Such datasets 
%requires a large number of %word targets 
%diversified input-to-text pairs 
 are generally not available in the neuroscience domain. Recurrent Neural Networks (RNNs) \citep{luong2015multi} decoders thereby have been traditionally more commonly implemented in existing works that address decoding tasks that involve neural activity. We therefore, as in prior works of brain-to-text decoding, adopt the two-stage system for brain-to-text decoding, where phonemes serve as the intermediate decoding targets. Notably, the diphone representation and the divide and conquer approach that we propose here do not rely on the particular architecture of the neural decoder and could the decoder could be interchanged without the need for adaptations. For existing benchmarks, such as Brain-to-Text 24', RNN neural decoder was identified as an optimal architecture for the neural decoder as we confirmed in our experiments in Table~\ref{table:model-architecture}. However, as further datasets become available the neural decoder is interchangeable with other potentially more optimal architectures for the neural decoder.

\paragraph{In-Context Learning}
LLMs pretrained on large corpora of texts exhibit the ability to learn new tasks in-context \citep{brown2020language}. That is, conditioning on a few demonstrations of input-target pairs, LLMs can generalize to unseen cases without updating their weights. This ICL ability has proven useful across a wide range of tasks \citep{wei2022chain, touvron2023llama}. While ICL typically underperforms a specialized LLM finetuned for a specific downstream task, it still surpasses zero-shot inference, and is particularly valuable when finetuning is not feasible due to resource constraints such as time or computational power, or the inacessibility of proprietary LLMs \citep{mosbach2023few}.

\section{Methods}
\label{Methods}
\begin{figure*}
\begin{center}
 \includegraphics[width=1\linewidth]{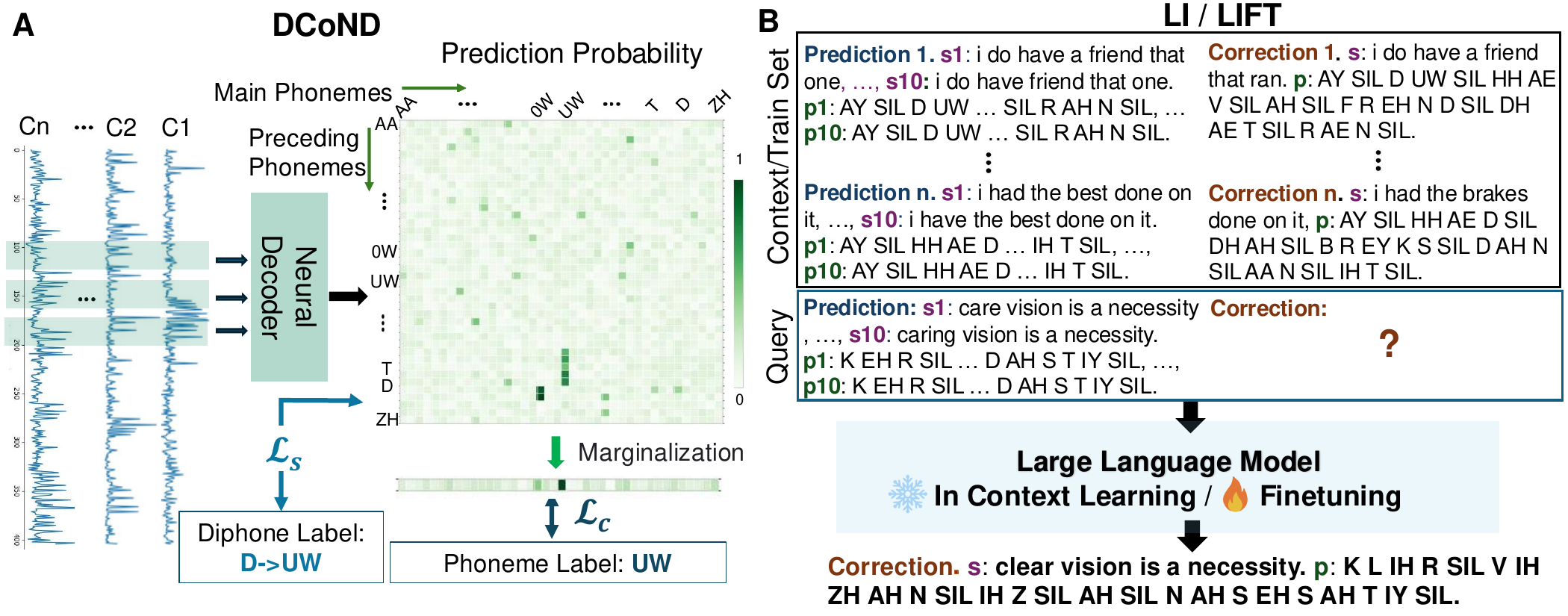}
\end{center}
   \caption{\textbf{A}: Illustration of the brain-to-phoneme decoding pipeline (DCoND). A Neural-Decoder in DCoND takes multi-channel neural signals as inputs and generates diphone probabilities, which are then marginalized into single phoneme probabilities.
   \textbf{B}: Illustration of the ensembling method for refining transcription predictions (LI/LIFT).  Given an ensemble of phoneme and transcription candidates as a query, GPT3.5 produces the most sensible transcription composed from these inputs. To do this, the LLM leverages examples of prediction-correction pairs provided either in-context at inference time (LI) or as training data during the finetuning process (LIFT).
 }
\label{fig:fig2-network-structure}
\end{figure*}

\paragraph{Problem formulation}
The problem of decoding phonemes from neural activity can be formulated as follows. Let $f : X \rightarrow Z$ be the mapping from neural activity $X \in \mathbb{R}^{T \times D} $ to 
phoneme sequence $Z\in \mathbb{Z}^{T^\prime}$, where $D$ is the number of neural features, $T$ is the number of neural time bins, 
and $T^\prime$ is the number of ground truth phonemes in a sentence. 
We note that $T > T^\prime$ in general, i.e. the articulation of one phoneme may span multiple timesteps. 
We also emphasize that there is no ground truth temporal alignment between $X$ and $Z$ due to the nature of the silent speech task. 
Both $T$ and $T^\prime$ vary across trials depending on the length of the sentence in that trial. We aim to learn a model $f_{\theta}: X \rightarrow Z$ to approximate $f$ with a set of parameters $\theta$, through a deep Neural-Decoder model. 

%We use an RNN model (GRU) for $f_\theta$. GRU has demonstrated superior performance on this dataset, as reported in previous works \citep{willett2023high, benchetrit2023brain}. 
For the Neural-Decoder, we employ an GRU recurrent neural network model. Previous work investigated various architectures for the Neural-Decoder and demonstrated that GRU shows superior performance \citep{willett2023high, benchetrit2023brain}.
A comparative study of alternative architectures, such as LSTM and Transformer, is available in the Appendix \ref{appendix:model-architecture}. 
In particular, we reconfirm the previous results of effectiveness of GRU vs. other architectures such as Transformers. Notably, the proposed divide-and-conquer strategy is adaptable to various decoder architecture and is designed to enhance the decoding performance of the any decoder that has been chosen.
Decoded phonemes $Z$ can be subsequently translated to sentences $Y$ with the help of a language model $h_\phi: Z \rightarrow Y$, where $h_\phi$ can be a pre-built statistical language model, e.g. 5-gram, or an LLM, e.g. GPT3 \citep{brown2020language}. The overall pipeline is depicted in Figure \ref{fig:fig1-task-illustraion}.

\paragraph{A Divide-and-Conquer strategy for phoneme decoding}
Decoding phonemes from neural activity is a nontrivial task given the highly nonlinear nature of $f$ and the variability of the neural population dynamics. Evidence exists that the neural representations for phonemes vary depending on the surrounding contexts \citep{bouchard2014neural, mugler2014direct}. We illustrate this observation in Figure \ref{fig:Fig4-representation-neuralspace} where segments of phoneme-aligned neural activity form clusters in the neural space based on the context they are in. It can be seen that there is no single cluster representing each phoneme, but rather each phoneme is represented by multiple subclusters. We further show that the subclusters are identifiable by the phoneme preceding the phoneme of interest. For instance, the phoneme AH is represented by subclusters DH $\rightarrow$ AH and SIL $\rightarrow$ AH (see further discussion in Section \ref{sec:phoneme-representation}). Learning to model these context-aware sub-units of speech instead of single phonemes directly could facilitate the phoneme decoding task. Concretely,
\begin{equation}
    f(x) := p(z|X) =  \sum_{S} p(z, S | X) = \sum_{s \in S} g_s^z(x)\\
    % P(Z=c_i|X) = \sum_{S}p(Z=c_i, S=c_j|X) = \sum_{c_j \in C}g_{c_j}^{c_i}(x)
\end{equation}\label{eq:subclasses expansion}
where $S$ is a random variable denoting the context surrounding the phoneme $z$. To be noticed, $z$ takes values from phoneme classes, such that $z \in [1, C]$. The problem of learning single phoneme classes ($f$) now reduces to the problem of learning the phoneme context-dependent subclasses ($g_s^z$), which is more manageable and in-line with the context-dependent nature of the data. We refer to our phoneme decoder with this divide-and-conquer strategy as \textbf{DCoND}.
% Here, we assume that the distribution of neural signal $X$ during a small time window is controlled by phoneme subclasses, including context information $s \in \mathcal{S}$ following a guassian distribution. Explicitly, we have:
% \begin{equation}
%     P(X|S=s_i) = \mathcal{G}(\bold{\mu}_{s_i}, \Sigma_{s_i}) %g_{s_i}(s_i)
% \end{equation}

% where $\mathcal{S}$ is the set of subclasses that extends the original phoneme set $\mathcal{C}$. During the process of generating context-dependent phoneme subset $s_i$, the neural distribution has mean $\mu_{s_i}$ covariance $ \Sigma_{s_i}$. For a phoneme class $c_i$, it constitutes of n subclasses, $c_i = \bigcup_{in-n:in} s_i$.
% To learn the probability of neural signal $X$ belongs to class $c_i$, we can collapse over the subclasses $s_i$:
% \begin{align}
%        P(C=c_i|X) &= \sum_{j \in [in-n:in]} P(S=s_j|X) 
% \end{align}\label{eq:expansion}
% %        \\  &= \sum_{j \in [in-n:in]} \frac{ g_{s_j}(s_j)P(s_j)}{P(X)}
% The idea of Expanding and Collapsing emphasizes that learning of each subclass $P(S=s_j|X)$, where the boundary of each subclass has more clear compared to the main class, would be easier than directly learning the distribution of the main class ($P(C=c_i|X)$).
% In addition, the collapsing of subclasses has additional benefits that allow model output to be robust to the noisy prediction of $P(S=s_i|X)$ where subclass samples are sparse.  We refer to our phoneme decoder with this divide-and-conquer strategy as \textbf{DCoND}.

\paragraph{Diphone as a context-dependent representation of phonemes} \label{sec:diphone collapse}
The context-dependent subclasses could be defined in multiple ways. In this work, we adopt diphone, a context-dependent representation for phoneme sequences where transitions between phonemes are the subject of interest. For example, the single phoneme representation of “hope”,
 $H,~~~OW,~~~P $,
will have a diphone representation
\[SIL\rightarrow H,~~~H\rightarrow H,~~~H\rightarrow OW,~~~OW\rightarrow OW,~~~OW \rightarrow P,~~~P\rightarrow P,~~~P\rightarrow SIL,\]
where ’SIL’ indicates the silence between the words. Diphone expands the length of phoneme sequence to $T^{\prime \prime}=2T^{\prime}$, where $T{\prime}$ and $T{\prime \prime}$ indicate the lengths of single phonemes and the diphone respectively, and increases the number of decoding classes to $C^2$, where the number of classes $C=40$ for the English language\footnote{the phonemes are defined as per CMU Pronouncing Dictionary: \url{http://www.speech.cs.cmu.edu/cgi-bin/cmudict/}}. 

Formally, we reformulate the problem of \textit{decoding a phoneme} from neural activity as the \textit{marginalization over the distribution of diphones}, conditioning on the observed neural activity
\[
%f(x) = 
p(Z=c_i|X) = \sum_{c_j\in S} p(c_j, c_i | X),
\]
where $p(c_j, c_i | X) $ is the probability of neural activity X encoding the diphone $c_j \rightarrow c_i $. A visualization of the marginalization process is shown in Fig. \ref{fig:fig2-network-structure}A. Neural activity is processed by a Neural Decoder to predict the probability of $40^2$ diphones being spoken at each timestep. The diphone probability is depicted by a $40 \times 40$ matrix where columns correspond to the current phonemes and rows correspond to the preceding phonemes. The single phoneme probability is then obtained by summing the joint probabilities column-wise.

\paragraph{Parameter Optimization for Phoneme Decoding}
As mentioned above, since there is no temporal alignment between $T$ timesteps of neural activity and $T'$ ground truth phonemes in each trial, we therefore use the Connectionist Temporal Classification (CTC) loss as proposed in \citep{graves2006connectionist} to resolve the unaligned sequences comparison. Specifically, for neural activity X and phoneme sequence Z, CTC aims to maximize the probability of Z given X
\begin{equation}
    p(Z|X) = \sum_{A \in \mathcal{A}_{(X,Z)}} \prod_{t=1}^{T'} p(a_{t}|X),
\end{equation}
where $A_{(X,Z)}$ is the set of valid alignments between $X$ and $Z$.

Given that we have the diphone representation for each ground truth sentence we consider the CTC losses over both the diphone and single phoneme representations
\begin{equation}
    \mathcal{L} = \alpha \mathcal{L}_c + (1-\alpha)\mathcal{L}_s,
    % -\alpha \log(\sum_{A \in \mathcal{A}_{(X,Z)}} \prod_{t=1}^{T'} p_m(a_{t}|X))  - (1-\alpha)\log(\sum_{A \in \mathcal{A}_{(X, C}} \prod_{t=1}^{2T'} p(a_{t}|X)),
\end{equation}\label{eq:loss}
where $\mathcal{L}_c = -\log(\sum_{A \in \mathcal{A}_{(X,Z)}} \prod_{t=1}^{T'} p_m(a_{t}|X))$ is the loss for single phoneme decoding, $\mathcal{L}_s  = -\log(\sum_{A \in \mathcal{A}_{(X, S)}} \prod_{t=1}^{T''} p(a_{t}|X))$ defines the loss over subclasses (diphone) decoding.

Coefficient $\alpha$ controls the balance of the single phoneme decoding and diphone decoding. $\alpha$ is designed to be small at the beginning and gradually increase over the course of training. See Appendix \ref{appendix:implementation detail} for more implementation details.

\paragraph{Word Decoding with Language Models}

The predicted phoneme probabilities are further transformed into high-quality text through (i) \textit{Generation} of transcription candidates from phonemes, (ii) \textit{Re-scoring} of transcription candidates, and (iii) \textit{Error correction} using an ensemble of selected candidates.

\textit{Transcription Generation.} During the phase of candidate sentence generation we convert the predicted phoneme probabilities into words using a 5-gram model. Based on the predicted phoneme probability distribution, the 5-gram model leverages its internal word and sentence distributions to generate the most likely sentence candidates~\citep{miao2015eesen,willett2023high}. Each candidate is associated with a likelihood score provided by the 5-gram model.
% The 5-gram model is trained on OpenWebText2 which has billions of words and a vocabulary size of 125,000.

\textit{Transcription Re-scoring} 
 LLMs trained on large corpora of texts, such as the Open Pre-trained Transformer (OPT) \citep{zhang2022opt}, could provide more accurate likelihood of the generated transcriptions. 
Hence, we use OPT to re-score the 5-gram likelihood outputs. The transcription candidates with the highest likelihoods are selected \citep{willett2023high}.

\textit{Transcription Error Correction with Ensemble Method} While the 5-gram and OPT models can correct some phoneme errors made by the phoneme decoder to produce more contextually sound sentences (transcriptions), these sentences are not always perfect. Variations of the phoneme decoding model could result in changes of generated and selected sentence candidates. Ensembles of phoneme decoding models, with each model being an expert in different situations, could mitigate the errors made by another model.

In \citep{benster2024cross} GPT3.5 is finetuned to evaluate an ensemble of $10$ transcription candidates and generate the most sensible sentence from the 10 candidates. However, providing GPT3.5 only the candidate transcriptions hinders the LLM's ability to understand the underlying phoneme sequences, which are the generating source of the transcriptions and might have been incorrectly converted by the 5-gram model. We therefore propose to include both the transcription candidates and the corresponding phoneme sequences as inputs to GPT3.5, tasking the model with generating both the correct transcription and phoneme sequence. An illustration of such a task is shown in Fig.\ref{fig:fig2-network-structure}. By finetuning the LLM in this manner, we train it to infer the relationship between predicted phonemes and the predicted transcriptions, as well as identifying common model-specific mistakes made by the phoneme decoders across their predictions. We show in Section \ref{sec:benchmark comparison} that this strategy further boosts the WER from 8.06\% to 5.77\%. 

In addition, since finetuning LLM is a resource-intensive process, we also propose to leverage ICL as an alternative learning paradigm for refining predicted transcriptions. Instead of finetuning GPT3.5 over multiple batches of ($10 \times$ predictions, $1 \times$ ground truth) pairs, we directly include $N$ examples of these pairs as context in each prompt, along with a query input to be refined. The LLM then leverages its ICL ability to quickly refine the query transcriptions without updating its weights. The prompts used for both in-context inference and finetuning are detailed in the Appendix \ref{appendix:gptprompt}.

\section{Experiments}
\subsection{Dataset}\label{sec:dataset}
We demonstrate the effectiveness of DCoND-LIFT in decoding attempted speech using the Brain-to-Text Benchmark 2024 \citep{willett2023high, willett2023highdataset}. The dataset was collected from a human subject with ALS who had lost the ability to produce intelligible speech. In the experiments, the subject attempts to silently speak sentences displayed on a screen. These sentences are composed from a vocabulary set of 125,000 words. In each trial, one sentence is shown followed by an auditory `Go' cue, after which the subject attempts to speak at their own pace. Neural activity (multiunit threshold crossings and spike band power) is recorded from the ventral premotor cortex (6V) while the subject attempted speaking. Due to the nature of the silent speech task, the correspondence between neural activity and the produced speech is unknown. The dataset is split into training, validation, and competition sets with 8800, 600, and 1200 sentences, respectively. 

%Focusing on phoneme error rate, demonstrating better phoneme error rate could be helpful for the word decoding as we l. The extending to the speech synthesis, perceptual word error rate. 

%The sound produced by ALS (Our participant (T12) has bulbar-onsetamyotrophic lateral sclerosis (ALS) patient is more similar to whisper The volume of whispered speech canvary depending on how forcefully air is pushed through, but is generally loudenough to be heard by other people who are nearby
\subsection{Evaluation Metrics}
\textbf{PER}  Phoneme Error Rate (PER) is calculated by comparing the decoded phoneme sequence with the ground truth phoneme sequence. After aligning the recognized phoneme sequence with the reference phoneme sequence, the number of insertions, deletions, and substitutions required to match the sequences are counted. The sum of these operations is divided by the total number of phonemes in the ground truth sequence to compute PER. This metric reflects how accurately neural signals can be recognized into phonetic units. 

\textbf{WER} Similarly to PER, word error rate (WER) is computed by aligning the sequence of recognized words with the ground truth sentence first and then counting the number of insertions, deletions, and substitutions of words needed to reconcile any discrepancies between the two sequences. The total number of these operations is divided by the total number of words in the reference sequence to obtain WER. As neural activity is translated into phonemes before converted into words, WER reflects the performance of both the neural decoder and the language model.

\textbf{P-WER} We adapt Perceptual Word Error Rate (P-WER) \citep{metzger2023high} to measure the quality of phoneme decoding at the word perception level. Specifically, we use eSpeak-NG \citep{espeakng}
%\footnote{\url{https://github.com/espeak-ng/espeak-ng}} 
to synthesize speech from the decoded phoneme sequences. Then the synthesized speech is translated into sentences by Whisper \citep{radford2022robust} from which the WER is estimated.
Considering the systematic errors introduced by the eSpeak-NG synthesizer and the Whisper ASR system, we define P-WER as follows
\[\text{P-WER}=(1 - \frac{1-\text{WER}_{\textit{Whisper-P}}}{1-\text{WER}_{\textit{Whisper-GT}}}), \]
where $\text{WER}_{Whisper-GT}$ and $\text{WER}_{Whisper-P}$ are the WER measured on Whisper's decoded transcriptions when audio is synthesized with ground truth phoneme sequences (GT) and predicted phoneme sequences (P), respectively.

\subsection{Comparison with SOTA Methods}\label{sec:benchmark comparison}
\begin{table}
  \caption{Performance comparison on Brain-to-Text 2024 Benchmark}
    \centering
  \label{table1:benchmark-comparison}
\begin{tabular}{lllll}
\toprule
 %& \cmidrule(r){1-3} \cmidrule(r){1-3}
          & PER$\times 100\downarrow$ & WER$\times 100\downarrow$ & P-WER$\times 100 \downarrow$ \\
\midrule
NPTL \citep{willett2023high} &    16.62  &  9.46   &  11.33                \\       
LISA \citep{benster2024cross}& ~~-- &  8.93  & ~~-- \\
\midrule

DCoND-L (Ours) &   \textbf{15.34}      &     8.06          &   \textbf{8.02 }         \\
%Tri-phone (5gram + OPT6B) &    15.11      &    9.81     &   31.35    &          6.38        \\
DCoND-LI (Ours) &    ~~--    &     7.29        &   ~~--        \\
DCoND-LIFT (Ours) &    ~~--      &    \textbf{5.77}          &  ~~--          \\
\bottomrule
\end{tabular}
\end{table}

Comparison of DCoND-LIFT performance with existing methods shows that DCoND-LIFT achieves state-of-the-art performance on the Brain-to-Text Benchmark 2024, where WER is the primary evaluation metric (see Table \ref{table1:benchmark-comparison}). Specifically, we compared DCoND-LIFT with the leading methods NPTL \citep{willett2023high} and LISA \citep{benster2024cross}. NPTL uses a 5-layer RNN to decode neural activity to phonemes, followed by a combination of 5-gram and OPT language models \citep{miao2015eesen, zhang2022opt} to translate decoded phonemes to text. 
%OPT is used to refine the 5-gram model outputs \citep{zhang2022opt}. %Generally, a 5-gram LM generates multiple candidates, each with corresponding scores indicating the possibility of each sentence. In the following step, the Open pre-trained transformer language model (OPT) \citep{zhang2022opt} is used to rescore the 5-gram outputs and select the best one based on these outputs and the prior knowledge of the OPT model.
LISA also uses RNN as phonemes decoder from neural activity, however it leverages GPT3.5 to further improve transcriptions given by the 5-gram model.
%The finetuning leveraging diversity from 10 different checkpoints. For each of the checkpoints, 5-gram model is leveraged to generate words. The 10 different versions of predicted sentences are further used to fine-tune ChatGPT. The fine-tuned model is then applied to the 5-gram outputs of the competition set to perform inference. 

As seen in Table \ref{table1:benchmark-comparison}, DCoND model variants outperform the competing methods. DCoND combined with 5-gram LM and OPT (DCoND-L) yields WER of 8.06\%, compared to 9.46\% WER of NPTL and 8.93\% of LISA. Further sensitivity analysis is provided in Table \ref{table:statistical_significance} of the Appendix. Given that DCoND-L uses the same backbone, RNN decoder and LMs, as NPTL, we posit that the improvements in WER are due to the effectiveness of our proposed divide-and-conquer phoneme decoding strategy. Indeed, DCoND-L achieves a better PER and P-WER (15.34\% and 8.02\% compared to 16.62\% and 11.33\% of NPTL). %, proving that modeling context-dependent phoneme representations facilitates the phoneme decoding task.

Further improvement in WER is achieved when we equip DCoND-L with the more powerful language model GPT3.5, to evaluate an ensemble of predicted transcriptions and their associated phoneme representations. When ensemble examplars are shown to GPT3.5 in-context (DCoND-LI), WER improves from 8.06\% to 7.29\%. This performance is achieved with 25 ICL examplars, the largest number of ICL examplars GPT3.5 can afford due to its prompt length constraint. When we finetune GPT3.5 with all available training examplars (DCoND-LIFT), WER is further boosted to 5.77\%, a significant improvement from 8.93\% WER of LISA. These results support our proposal of including both transcriptions and phoneme representations in the demonstrations to GPT3.5 so that it can leverage the relationship between phonemes and words to refine the transcriptions. 

\subsection{Phoneme Decoding Analyses}
\begin{figure}
\begin{center}
 \includegraphics[width=0.95\linewidth]{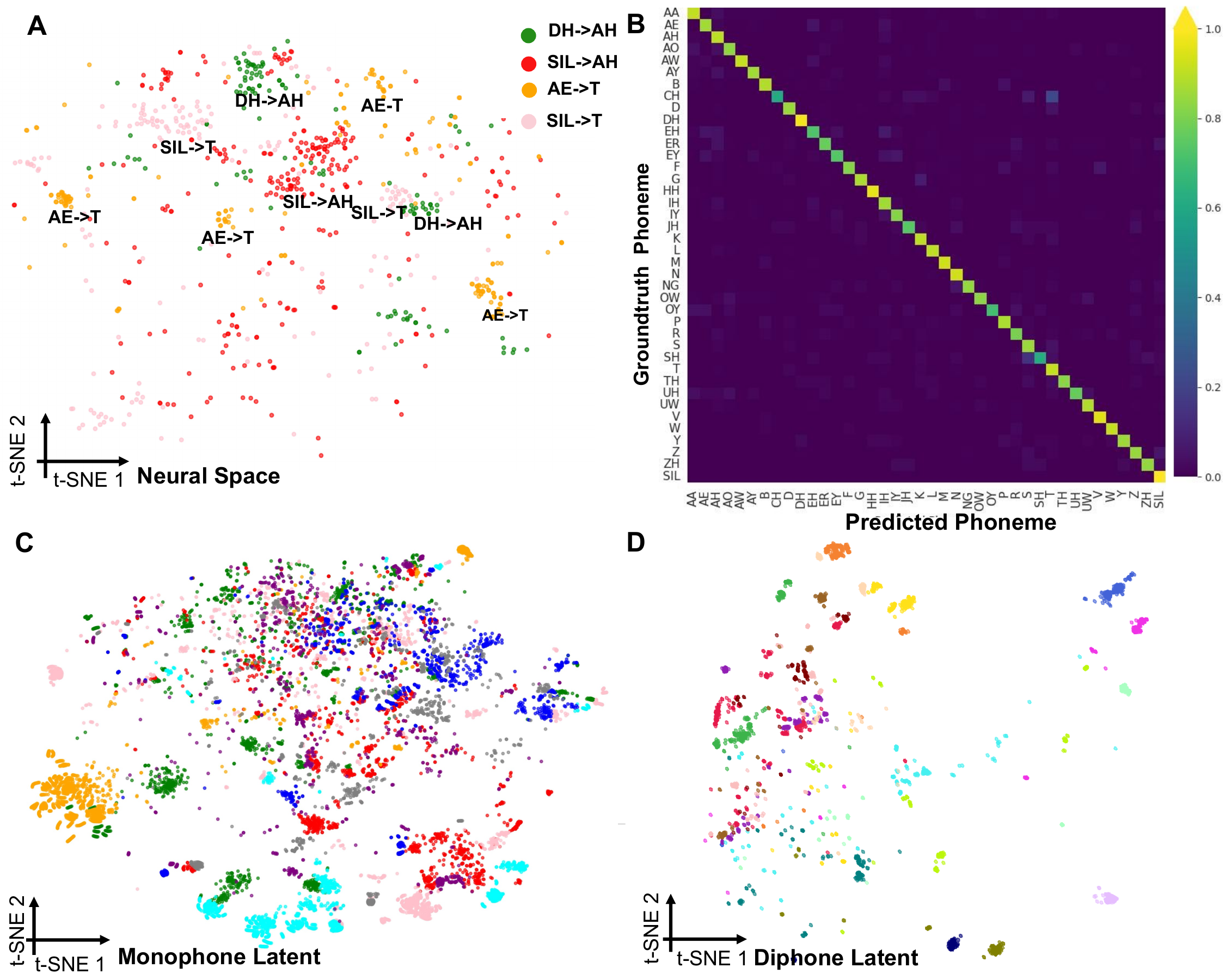}
\end{center}
   \caption{\textbf{A}: 2D t-SNE visualization of neural signal projections  illustrating the context-dependent nature of phonemes in neural reprentations. Different colors indicate different diphone classes. \textbf{B}: Confusion matrix of ground truth phonemes vs. DCoND's predicted phonemes. \textbf{C}: 2D t-SNE visualization for the latent space of the neural decoder trained with single phoneme decoding objective (Monophone). Different colors indicate different phoneme classes. \textbf{D}: 2D t-SNE visualization for the latent space of the neural decoder trained with diphone decoding objective. Different colors indicate different diphone classes. 
 }
\label{fig:Fig4-representation-neuralspace}
\end{figure}

\paragraph{Neural activity represents phonemes in context-dependent clusters}\label{sec:phoneme-representation}
%One fundamental question is how the neural signal encodes and generates different sounds. Since sound is a continuous sequence, matching neural signals to speech can be complicated. By using the monophone or transphone definitions of speech, we simplify the problem from a continuous-to-continuous mapping into a continuous-to-discrete mapping.
Previous works demonstrate that the decoding accuracy of phonemes from neural activity could be reduced when phonemes are pronounced in the context of other phonemes as opposed to being pronounced individually \citep{mugler2014direct}. To get a glimpse of how the brain encodes phonemes, in Fig. \ref{fig:Fig4-representation-neuralspace}A we visualize phoneme-aligned segments of neural activity in the 2D t-SNE space \citep{JMLR:v9:vandermaaten08a}. 
 %how neural signals are organized in the low-dimensional latent space and illustrate how neural coding of phonemes could be context-dependent. 
 %identify their properties related to the phoneme they produce. 
 Since the dataset does not have the exact temporal correspondence between neural activity and phonemes, we leverage Dynamic Time Warping (DTW) to align the ground truth phonemes to neural activity segments according to the timestamps obtained from the decoded phonemes \citep{muller2007dynamic}. We annotate the neural activity segments based on the resulting phoneme alignment.
The visualization reveals that neural activity segments form distinct clusters in the t-SNE space. Notably, these clusters are organized based not only on single phonemes but also on the context in which they are spoken. For instance, during periods where `T' is the main phoneme being spoken, corresponding neural activity is organized into subclusters of AE$\rightarrow$T (orange) and SIL$\rightarrow$T (pink), depending on whether phoneme `AE' or `SIL' is spoken before `T'. Similar observations hold for subclusters DH$\rightarrow$AH (green) and SIL$\rightarrow$AH (red) for the phoneme `AH'. We note that further subclusters could exist within each subcluster, suggesting a continuum of finer contexts beyond the preceding phoneme. 
% For example, for transitions ending with the 'T' sound, during different transitions like AE->T' and SIL->T', the neural signals form distinct clusters (orange cluster vs. pink cluster). Similarly, for DH->AH' (green) vs. SIL->AH' (red). Within the 'AE->T' transition, neural signals form subclusters, indicating that neural signals could depend on more context when generating speech signals.

\textbf{Diphone decoding leads to enhanced clusters in latent space} We visualize in  Figure \ref{fig:Fig4-representation-neuralspace}C and \ref{fig:Fig4-representation-neuralspace}D the latent space at the last layer of the neural decoder when trained to decode single phonemes (monophones) vs. diphones. In Figure \ref{fig:Fig4-representation-neuralspace}C, each color represents a single decoded phoneme label. For clarity of the visualization, we selected five single phoneme classes with the most samples. The clusters that correspond to single phonemes appear to spread out over the whole space, and overlap with each other.
In Figure \ref{fig:Fig4-representation-neuralspace}D, each color represents a decoded diphone. Since there are fewer samples for each diphone, we visualize 16 diphone classes with the highest occurrence. It can be observed that the neural decoder represents diphones in the latent space by clusters that are significantly more condensed and well-separated. Such clear structure facilitates the subsequent classification of single phonemes and demonstrates the effectiveness of our divide-and-conquer phoneme decoding method.

\textbf{Phoneme prediction error analysis} In Figure \ref{fig:Fig4-representation-neuralspace}B, we show the confusion matrix of the predicted phonemes and the ground truth phonemes. It can be seen from the figure that most phonemes are correctly classified with accuracy greater than 80\%. The mistakes that the model typically makes, if any, are on phonemes that are pronounced similarly.
For example, the model usually confuses `SH' with `S', and `CH' with `TH'. Since the articulation of these phonemes is very similar, the neural activity generating them is likely to be similar. Such confusion is expected to some extent, given the ALS condition hindering the ability of the subject to articulate the desired words clearly.

\subsection{Ablation Study}

\paragraph{Trade-off Between Diphone Loss and Monophone Loss}
We systematically investigate the trade-off between diphone loss $\mathcal{L}_c$ and monophone loss $\mathcal{L}_s$, controlled by the parameter $\alpha$ in Equation \ref{eq:loss}. The impact of varying $\alpha$ on model performance is shown in Table \ref{table:ablation-alpha}. We find that a balance between these two losses, with $\alpha = 0.6$, yields the most optimal results. Consequently, we adopt $\alpha = 0.6$ for all DCoND models used in this paper.

\begin{table}
    \caption{Trade-offs between diphone loss and monophone loss.}
    \label{table:ablation-alpha}
    \centering
    \begin{tabular}{lllllll}
    \toprule
     & $\alpha=0.2$   & $\alpha=0.4$ & $\alpha=0.6$  & $\alpha=0.8$ & $\alpha=1.0$ \\
    &&&(DCoND-L)& &(NPTL) \\
    \midrule
    PER$\times 100\downarrow$ & 15.64  & 15.26 &      15.34           &  15.49          & 16.62                   \\
    WER$\times 100\downarrow$ &8.47     & 8.70  &  8.06 & 8.64 & 9.46             \\
    \bottomrule
    \end{tabular}
\end{table} %& 22.48
\paragraph{Alternatives for context-dependent phoneme representations}
\begin{table*}
  \caption{Ablation study on alternative definitions of context-aware phoneme representations.}
  \label{table:alternativesubclass}
    \centering
  \adjustbox{max width=\textwidth}{
\begin{tabular}{lllllll}
\toprule
 %&& & \cmidrule(l){4} \\
 & & \multicolumn{4}{c}{Triphone} \\
\cmidrule(l){3-6}
         &  DCoND-L &		K=50&	K=100	&K=200	&Grouping   \\
\midrule
PER$\times 100\downarrow$ &  15.34	&	16.01	&\textbf{15.02}&	15.11	&28.55 \\
WER$\times 100\downarrow$ & \textbf{8.06}		&9.69&	9.67	&9.81	&13.98 \\
%P-WER$\times 100\downarrow$ &8.02 &	7.32		&6.6	&\textbf{6.38} &	52.12	\\
% wer 5gram \\
\bottomrule
\end{tabular}}
\end{table*}

Besides diphone, triphone is another way to define context-dependent representations for phonemes. Each triphone class consists of three consecutive phonemes, e.g. $H\rightarrow OW\rightarrow P$, providing a finer granularity of context dependency with $40^3$ possible classes. Such a large number of classes can be overwhelming for the model to learn. Given that many of them have few to no presence in the data, to efficiently maintain a manageable size of decoding classes we select the top $K$ combinations of preceding and succeeding phonemes for each main phoneme, e.g. $*\rightarrow OW\rightarrow *$, based on their frequency of occurrence in the data, where $K \in [50, 100, 200]$. Alternatively, the preceding and succeeding phonemes could be grouped based on their articulatory similarity ("Grouping" in Table \ref{table:alternativesubclass}) (see Appendix \ref{appendex:triphone} for more details).

Results in Table \ref{table:alternativesubclass} suggest that triphone with appropriate class size achieves comparable PER as the diphone counterpart (DCoND-L). However, triphone modeling underperforms diphone modeling in terms of WER, possibly due to the reduction in the triphone class size that is potentially skewing the phoneme distribution output of the neural decoder, making it incompatible with the distribution the subsequent 5-gram model was originally trained on single phoneme. Notably, while the grouping method despite yields a class size similar to that of $K=200$, it performs signficantly worse in both PER and WER. This implies that neural encoding of phonemes is more intricate, and grouping phonemes based on pronunciation similarity may not be optimal. Overall, we empirically find that diphone, with its context-dependent nature and manageable class size, as the most suitable modeling choice for this task and dataset.
\begin{wrapfigure}{r}{0.5\textwidth}
%\begin{figure*}
\begin{center}
 \includegraphics[width=1\linewidth]{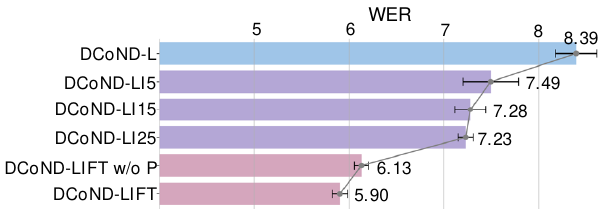}
\end{center}
   \caption{Ablation study on the contribution of LLMs.
 }
\label{fig:Fig5-wer-ablation}
%\end{figure*}
\end{wrapfigure}

\paragraph{Contribution of LLMs}
LLMs play an important role in translating phonemes into sentences. As detailed in Section~\ref{Methods}, our LLM-based phoneme-to-text pipeline consists of three steps: (i) transcription generation (\textit{5-gram}), (ii) transcription re-scoring (\textit{OPT}), (iii) error correction via ensemble with \textit{ICL GPT3.5} or \textit{finetuned GPT3.5}. We show in Figure~\ref{fig:Fig5-wer-ablation} how each step of the LLM pipeline contributes to the overall WER. 
In particular, we consider the following variants of LLMs, as an added component to DCoND: \textit{5-gram}+\textit{OPT} as used in NPTL (DCoND-L), \textit{5-gram}+\textit{OPT}+\textit{ICL GPT3.5} with context length of 5 (DCoND-LI5), context length of 15 (DCoND-LI15) and context length of 25 (DCoND-LI25), \textit{5-gram}+\textit{OPT}+\textit{finetuned GPT3.5} without phoneme inputs (DCoND-LIFT w/o P), and our most performative model -- \textit{5-gram}+\textit{OPT}+\textit{finetuned GPT3.5} with phoneme inputs (DCoND-LIFT). We show that the use of GPT3.5 to refine the transcriptions from an ensemble of candidates, selected from the highest re-scored likelihood given by the \textit{5-gram}+\textit{OPT} step, leads to an improvement in WER.  The most optimal WER is achieved when GPT3.5 fully leverages the predicted phonemes to refine query transcriptions (DCoND-LIFT). In an ablated version, where limited ICL exemplars are given to GPT3.5 without fine-tuning (DCoND-LIxx), the WER score is enhanced when compared to the baseline (DCoND-L), but does not fully reach the accuracy of DCoND-LIFT. Such disparity between the two modes of LLM effectiveness is expected since the two modes are designed for different operational regimes: DCoND-LI is applicable in scenarios with fast inference and constraints on response time, while DCoND-LIFT is applicable in scenarios in which such constraints do not exist. Additional ablations are provided in Section \ref{appendix:ablation_llm} of the Appendix.
% Generally without ChatGPT enhancement ( 5gram, 5gram+OPT), the performance is lower than directly using ChatGPT with ICL without any fine-tuning (ECoLM-I). The ChatGPT learns the correction from the provided context and the WER decreases as more context is provided.  ECoLM-F with the fine-tuning equivalent as taking a much longer context, achieves better performance. Especially when decoded phoneme sequences are provided as part of the inputs.
\section{Conclusion}\label{sec:discussion}
In this work, we propose a Divide-and-Conquer approach for Neural Decoders (DCoND) together with an LLM-enhanced ensemble method (LI and LIFT) for speech decoding from neural activity. DCoND is motivated by the coarticulation principle in neuroscience research which informs that there is context-dependent organization of phonemes in the brain.  In particular, DCoND proposes to utilize diphones, a context-dependent representation for phoneme sequences, as the modeling target. We show that decomposing the phoneme classification task into diphone classification subtasks facilitates the phoneme decoding task, subsequently improving the whole sentence decoding accuracy. Furthermore, variants of DCoND (DCoND-LI and DCoND-LIFT) propose the additional inclusion of an LLM-based ensemble approach to enhance DCoND ability to refine the transcription candidates. In particular, these approaches leverage GPT3.5 in two modes, ICL and fine-tuning, to further map between phoneme sequence candidates and transcription candidates and thus achieve further improvement in contextual word decoding.  We show that DCoND-LIFT achieves SOTA PER and WER on the Brain-to-Text 2024 Benchmark, outperforming leading methods by a large margin. 
% The proposed method improves the neural decoding for more accurately estimating the acoustic unit corresponding to the neural signal and the LLM-enhanced correction on the decoded sentence outputs. DCoND divides the neural phoneme decoding problem into subproblems that are more aligned with the neural coding mechanism to enhance the decoding accuracy. LIFT applies ChatGPT to correct the generated text from 5-gram+OPT outputs, demonstrating ChatGPT could learn to correct the generated sentences using correction examples without any fine-tuning while fine-tuning equivalently to using longer context achieves better performance. 
% With DCoND and LIFT, the decoding accuracy on the Brain-to-Text benchmark 2024 improves from the 8.96\% to 5.81\%. 

% \paragraph{Limitations}
% The current decoding pipeline involves two stages: decoding  phonemes from neural activity and converting the decoded phonemes into texts. The decoding of phonemes using the triphone representation shows a lower PER but higher WER. This could be due to the process of selecting top K triphone subclasses making the phoneme distribution inconsistent with language models. A better method of choosing the subset of K triphones could be developed to further enhance the decoding performance. 

\section{Acknowledgements}
We acknowledge the support of HDR Institute: Accelerated
AI Algorithms for Data-Driven Discovery (A3D3) National
 Science Foundation grant PHY-2117997 (ES, JL, TL, MC) and the departments of Applied Mathematics (ES) and Electrical and Computer Engineering (ES, JL, TL, MC) at the University of Washington. The  authors are thankful to the Center of Computational Neuroscience and the eScience Center at the University of Washington. 

\bibliography{iclr2025_conference}

\begin{thebibliography}{54}
\providecommand{\natexlab}[1]{#1}
\providecommand{\url}[1]{\texttt{#1}}
\expandafter\ifx\csname urlstyle\endcsname\relax
  \providecommand{\doi}[1]{doi: #1}\else
  \providecommand{\doi}{doi: \begingroup \urlstyle{rm}\Url}\fi

\bibitem[Aggarwal \& Dave(2011)Aggarwal and Dave]{aggarwal2011acoustic}
Rajesh~Kumar Aggarwal and Mayank Dave.
\newblock Acoustic modeling problem for automatic speech recognition system: conventional methods (part i).
\newblock \emph{International Journal of Speech Technology}, 14:\penalty0 297--308, 2011.

\bibitem[Benchetrit et~al.(2023)Benchetrit, Banville, and King]{benchetrit2023brain}
Yohann Benchetrit, Hubert Banville, and Jean-R{\'e}mi King.
\newblock Brain decoding: toward real-time reconstruction of visual perception.
\newblock \emph{arXiv preprint arXiv:2310.19812}, 2023.

\bibitem[Benster et~al.(2024)Benster, Wilson, Elisha, Willett, and Druckmann]{benster2024cross}
Tyler Benster, Guy Wilson, Reshef Elisha, Francis~R Willett, and Shaul Druckmann.
\newblock A cross-modal approach to silent speech with llm-enhanced recognition.
\newblock \emph{arXiv preprint arXiv:2403.05583}, 2024.

\bibitem[Bouchard \& Chang(2014)Bouchard and Chang]{bouchard2014neural}
Kristofer~E Bouchard and Edward~F Chang.
\newblock Neural decoding of spoken vowels from human sensory-motor cortex with high-density electrocorticography.
\newblock In \emph{2014 36th Annual International Conference of the IEEE Engineering in Medicine and Biology Society}, pp.\  6782--6785. IEEE, 2014.

\bibitem[Brown et~al.(2020)Brown, Mann, Ryder, Subbiah, Kaplan, Dhariwal, Neelakantan, Shyam, Sastry, Askell, Agarwal, Herbert-Voss, Krueger, Henighan, Child, Ramesh, Ziegler, Wu, Winter, Hesse, Chen, Sigler, Litwin, Gray, Chess, Clark, Berner, McCandlish, Radford, Sutskever, and Amodei]{brown2020language}
Tom~B Brown, Benjamin Mann, Nick Ryder, Melanie Subbiah, Jared Kaplan, Prafulla Dhariwal, Arvind Neelakantan, Pranav Shyam, Girish Sastry, Amanda Askell, Sandhini Agarwal, Ariel Herbert-Voss, Gretchen Krueger, Tom Henighan, Rewon Child, Aditya Ramesh, Daniel~M Ziegler, Jeffrey Wu, Clemens Winter, Christopher Hesse, Mark Chen, Eric Sigler, Mateusz Litwin, Scott Gray, Benjamin Chess, Jack Clark, Christopher Berner, Sam McCandlish, Alec Radford, Ilya Sutskever, and Dario Amodei.
\newblock Language models are few-shot learners.
\newblock \emph{Advances in neural information processing systems}, 33:\penalty0 1877--1901, 2020.

\bibitem[Card et~al.(2024)Card, Wairagkar, Iacobacci, Hou, Singer-Clark, Willett, Kunz, Fan, Nia, Deo, Srinivasan, Choi, Glasser, Hochberg, Henderson, Shahlaie, Stavisky, and Brandman]{card2024accurate}
Nicholas~S. Card, Maitreyee Wairagkar, Carrina Iacobacci, Xianda Hou, Tyler Singer-Clark, Francis~R. Willett, Erin~M. Kunz, Chaofei Fan, Maryam~Vahdati Nia, Darrel~R. Deo, Aparna Srinivasan, Eun~Young Choi, Matthew~F. Glasser, Leigh~R. Hochberg, Jaimie~M. Henderson, Kiarash Shahlaie, Sergey~D. Stavisky, and David~M. Brandman.
\newblock An accurate and rapidly calibrating speech neuroprosthesis.
\newblock \emph{New England Journal of Medicine}, 391\penalty0 (7):\penalty0 609--618, 2024.

\bibitem[Chen et~al.(2024)Chen, Wang, Khalilian-Gourtani, Yu, Dugan, Friedman, Doyle, Devinsky, Wang, and Flinker]{chen2024neural}
Xupeng Chen, Ran Wang, Amirhossein Khalilian-Gourtani, Leyao Yu, Patricia Dugan, Daniel Friedman, Werner Doyle, Orrin Devinsky, Yao Wang, and Adeen Flinker.
\newblock A neural speech decoding framework leveraging deep learning and speech synthesis.
\newblock \emph{Nature Machine Intelligence}, pp.\  1--14, 2024.

\bibitem[Darjaa et~al.(2011)Darjaa, Cer{\v{n}}ak, Be{\v{n}}u{\v{s}}, Rusko, Sabo, and Trnka]{darjaa2011rule}
Sakhia Darjaa, Milo{\v{s}} Cer{\v{n}}ak, {\v{S}}tefan Be{\v{n}}u{\v{s}}, Milan Rusko, R{\'o}bert Sabo, and Mari{\'a}n Trnka.
\newblock Rule-based triphone mapping for acoustic modeling in automatic speech recognition.
\newblock In \emph{Text, Speech and Dialogue: 14th International Conference, TSD 2011, Pilsen, Czech Republic, September 1-5, 2011. Proceedings 14}, pp.\  268--275. Springer, 2011.

\bibitem[D{\'e}fossez et~al.(2023)D{\'e}fossez, Caucheteux, Rapin, Kabeli, and King]{defossez2023decoding}
Alexandre D{\'e}fossez, Charlotte Caucheteux, J{\'e}r{\'e}my Rapin, Ori Kabeli, and Jean-R{\'e}mi King.
\newblock Decoding speech perception from non-invasive brain recordings.
\newblock \emph{Nature Machine Intelligence}, 5\penalty0 (10):\penalty0 1097--1107, 2023.

\bibitem[Diener et~al.(2018)Diener, Felsch, Angrick, and Schultz]{diener2018session}
Lorenz Diener, Gerrit Felsch, Miguel Angrick, and Tanja Schultz.
\newblock Session-independent array-based emg-to-speech conversion using convolutional neural networks.
\newblock In \emph{Speech Communication; 13th ITG-Symposium}, pp.\  1--5. VDE, 2018.

\bibitem[Dong et~al.(2018)Dong, Xu, and Xu]{dong2018speech}
Linhao Dong, Shuang Xu, and Bo~Xu.
\newblock Speech-transformer: a no-recurrence sequence-to-sequence model for speech recognition.
\newblock In \emph{2018 IEEE international conference on acoustics, speech and signal processing (ICASSP)}, pp.\  5884--5888. IEEE, 2018.

\bibitem[Fodor et~al.(2024)Fodor, Csap{\'o}, and Arthur]{fodor2024towards}
Mil{\'a}n~Andr{\'a}s Fodor, Tam{\'a}s~G{\'a}bor Csap{\'o}, and Frigyes~Viktor Arthur.
\newblock Towards decoding brain activity during passive listening of speech.
\newblock \emph{arXiv preprint arXiv:2402.16996}, 2024.

\bibitem[Gaddy \& Klein(2020)Gaddy and Klein]{gaddy2020digital}
David Gaddy and Dan Klein.
\newblock Digital voicing of silent speech.
\newblock \emph{arXiv preprint arXiv:2010.02960}, 2020.

\bibitem[Gaddy \& Klein(2021)Gaddy and Klein]{gaddy2021improved}
David Gaddy and Dan Klein.
\newblock An improved model for voicing silent speech.
\newblock \emph{arXiv preprint arXiv:2106.01933}, 2021.

\bibitem[Graves(2012)]{graves2012sequence}
Alex Graves.
\newblock Sequence transduction with recurrent neural networks.
\newblock \emph{arXiv preprint arXiv:1211.3711}, 2012.

\bibitem[Graves et~al.(2006)Graves, Fern{\'a}ndez, Gomez, and Schmidhuber]{graves2006connectionist}
Alex Graves, Santiago Fern{\'a}ndez, Faustino Gomez, and J{\"u}rgen Schmidhuber.
\newblock Connectionist temporal classification: labelling unsegmented sequence data with recurrent neural networks.
\newblock In \emph{Proceedings of the 23rd international conference on Machine learning}, pp.\  369--376, 2006.

\bibitem[Gulati et~al.(2020)Gulati, Qin, Chiu, Parmar, Zhang, Yu, Han, Wang, Zhang, Wu, and Pang]{gulati2020conformer}
Anmol Gulati, James Qin, Chung-Cheng Chiu, Niki Parmar, Yu~Zhang, Jiahui Yu, Wei Han, Shibo Wang, Zhengdong Zhang, Yonghui Wu, and Ruoming Pang.
\newblock Conformer: Convolution-augmented transformer for speech recognition.
\newblock \emph{arXiv preprint arXiv:2005.08100}, 2020.

\bibitem[Herff et~al.(2015)Herff, Heger, De~Pesters, Telaar, Brunner, Schalk, and Schultz]{herff2015brain}
Christian Herff, Dominic Heger, Adriana De~Pesters, Dominic Telaar, Peter Brunner, Gerwin Schalk, and Tanja Schultz.
\newblock Brain-to-text: decoding spoken phrases from phone representations in the brain.
\newblock \emph{Frontiers in neuroscience}, 9:\penalty0 217, 2015.

\bibitem[Hsu et~al.(2021)Hsu, Bolte, Tsai, Lakhotia, Salakhutdinov, and Mohamed]{hsu2021hubert}
Wei-Ning Hsu, Benjamin Bolte, Yao-Hung~Hubert Tsai, Kushal Lakhotia, Ruslan Salakhutdinov, and Abdelrahman Mohamed.
\newblock Hubert: Self-supervised speech representation learning by masked prediction of hidden units.
\newblock \emph{IEEE/ACM Transactions on Audio, Speech, and Language Processing}, 29:\penalty0 3451--3460, 2021.

\bibitem[Huang et~al.(2014)Huang, Baker, and Reddy]{huang2014historical}
Xuedong Huang, James Baker, and Raj Reddy.
\newblock A historical perspective of speech recognition.
\newblock \emph{Communications of the ACM}, 57\penalty0 (1):\penalty0 94--103, 2014.

\bibitem[Janke \& Diener(2017)Janke and Diener]{janke2017emg}
Matthias Janke and Lorenz Diener.
\newblock Emg-to-speech: Direct generation of speech from facial electromyographic signals.
\newblock \emph{IEEE/ACM Transactions on Audio, Speech, and Language Processing}, 25\penalty0 (12):\penalty0 2375--2385, 2017.

\bibitem[Jou et~al.(2006)Jou, Schultz, Walliczek, Kraft, and Waibel]{jou2006towards}
Szu-Chen Jou, Tanja Schultz, Matthias Walliczek, Florian Kraft, and Alex Waibel.
\newblock Towards continuous speech recognition using surface electromyography.
\newblock In \emph{Ninth International Conference on Spoken Language Processing}, 2006.

\bibitem[Kapur et~al.(2018)Kapur, Kapur, and Maes]{kapur2018alterego}
Arnav Kapur, Shreyas Kapur, and Pattie Maes.
\newblock Alterego: A personalized wearable silent speech interface.
\newblock In \emph{23rd International conference on intelligent user interfaces}, pp.\  43--53, 2018.

\bibitem[Kellis et~al.(2010)Kellis, Miller, Thomson, Brown, House, and Greger]{kellis2010decoding}
Spencer Kellis, Kai Miller, Kyle Thomson, Richard Brown, Paul House, and Bradley Greger.
\newblock Decoding spoken words using local field potentials recorded from the cortical surface.
\newblock \emph{Journal of neural engineering}, 7\penalty0 (5):\penalty0 056007, 2010.

\bibitem[LAleye et~al.(2016)LAleye, Besacier, Ezin, and Motamed]{laleye2016first}
Fr{\'e}jus~AA LAleye, Laurent Besacier, Eug{\`e}ne~C Ezin, and Cina Motamed.
\newblock First automatic fongbe continuous speech recognition system: Development of acoustic models and language models.
\newblock In \emph{2016 Federated Conference on Computer Science and Information Systems (FedCSIS)}, pp.\  477--482. IEEE, 2016.

\bibitem[Linse et~al.(2018)Linse, Aust, Joos, and Hermann]{linse2018communication}
Katharina Linse, Elisa Aust, Markus Joos, and Andreas Hermann.
\newblock Communication matters—pitfalls and promise of hightech communication devices in palliative care of severely physically disabled patients with amyotrophic lateral sclerosis.
\newblock \emph{Frontiers in neurology}, 9:\penalty0 379945, 2018.

\bibitem[Luong et~al.(2015)Luong, Le, Sutskever, Vinyals, and Kaiser]{luong2015multi}
Minh-Thang Luong, Quoc~V Le, Ilya Sutskever, Oriol Vinyals, and Lukasz Kaiser.
\newblock Multi-task sequence to sequence learning.
\newblock \emph{arXiv preprint arXiv:1511.06114}, 2015.

\bibitem[Meltzner et~al.(2018)Meltzner, Heaton, Deng, De~Luca, Roy, and Kline]{meltzner2018development}
Geoffrey~S Meltzner, James~T Heaton, Yunbin Deng, Gianluca De~Luca, Serge~H Roy, and Joshua~C Kline.
\newblock Development of semg sensors and algorithms for silent speech recognition.
\newblock \emph{Journal of neural engineering}, 15\penalty0 (4):\penalty0 046031, 2018.

\bibitem[Metzger et~al.(2022)Metzger, Liu, Moses, Dougherty, Seaton, Littlejohn, Chartier, Anumanchipalli, Tu-Chan, Ganguly, and Chang]{metzger2022generalizable}
Sean~L. Metzger, Jessie~R. Liu, David~A. Moses, Maximilian~E. Dougherty, Margaret~P. Seaton, Kaylo~T. Littlejohn, Josh Chartier, Gopala~K. Anumanchipalli, Adelyn Tu-Chan, Karunesh Ganguly, and Edward~F. Chang.
\newblock Generalizable spelling using a speech neuroprosthesis in an individual with severe limb and vocal paralysis.
\newblock \emph{Nature communications}, 13\penalty0 (1):\penalty0 6510, 2022.

\bibitem[Metzger et~al.(2023)Metzger, Littlejohn, Silva, Moses, Seaton, Wang, Dougherty, Liu, Wu, Berger, Zhuravleva, Tu-Chan, Ganguly, Anumanchipalli, and Chang]{metzger2023high}
Sean~L. Metzger, Kaylo~T. Littlejohn, Alexander~B. Silva, David~A. Moses, Margaret~P. Seaton, Ran Wang, Maximilian~E. Dougherty, Jessie~R. Liu, Peter Wu, Michael~A. Berger, Inga Zhuravleva, Adelyn Tu-Chan, Karunesh Ganguly, Gopala~K. Anumanchipalli, and Edward~F. Chang.
\newblock A high-performance neuroprosthesis for speech decoding and avatar control.
\newblock \emph{Nature}, pp.\  1--10, 2023.

\bibitem[Miao et~al.(2015)Miao, Gowayyed, and Metze]{miao2015eesen}
Yajie Miao, Mohammad Gowayyed, and Florian Metze.
\newblock Eesen: End-to-end speech recognition using deep rnn models and wfst-based decoding.
\newblock In \emph{2015 IEEE workshop on automatic speech recognition and understanding (ASRU)}, pp.\  167--174. IEEE, 2015.

\bibitem[Mosbach et~al.(2023)Mosbach, Pimentel, Ravfogel, Klakow, and Elazar]{mosbach2023few}
Marius Mosbach, Tiago Pimentel, Shauli Ravfogel, Dietrich Klakow, and Yanai Elazar.
\newblock Few-shot fine-tuning vs. in-context learning: A fair comparison and evaluation.
\newblock \emph{arXiv preprint arXiv:2305.16938}, 2023.

\bibitem[Moses et~al.(2021)Moses, Metzger, Liu, Anumanchipalli, Makin, Sun, Chartier, Dougherty, Liu, Abrams, Tu-Chan, Ganguly, and Chang]{moses2021neuroprosthesis}
David~A. Moses, Sean~L. Metzger, Jessie~R. Liu, Gopala~K. Anumanchipalli, Joseph~G. Makin, Pengfei~F. Sun, Josh Chartier, Maximilian~E. Dougherty, Patricia~M. Liu, Gary~M. Abrams, Adelyn Tu-Chan, Karunesh Ganguly, and Edward~F. Chang.
\newblock Neuroprosthesis for decoding speech in a paralyzed person with anarthria.
\newblock \emph{New England Journal of Medicine}, 385\penalty0 (3):\penalty0 217--227, 2021.

\bibitem[Mugler et~al.(2014)Mugler, Patton, Flint, Wright, Schuele, Rosenow, Shih, Krusienski, and Slutzky]{mugler2014direct}
Emily~M Mugler, James~L Patton, Robert~D Flint, Zachary~A Wright, Stephan~U Schuele, Joshua Rosenow, Jerry~J Shih, Dean~J Krusienski, and Marc~W Slutzky.
\newblock Direct classification of all american english phonemes using signals from functional speech motor cortex.
\newblock \emph{Journal of neural engineering}, 11\penalty0 (3):\penalty0 035015, 2014.

\bibitem[M{\"u}ller(2007)]{muller2007dynamic}
Meinard M{\"u}ller.
\newblock Dynamic time warping.
\newblock \emph{Information retrieval for music and motion}, pp.\  69--84, 2007.

\bibitem[Nedel et~al.(2000)Nedel, Singh, and Stern]{nedel2000phone}
Jon~P Nedel, Rita Singh, and Richard~M Stern.
\newblock Phone transition acoustic modeling: application to speaker independent and spontaneous speech systems.
\newblock In \emph{INTERSPEECH}, pp.\  572--575, 2000.

\bibitem[Pandarinath et~al.(2017)Pandarinath, Nuyujukian, Blabe, Sorice, Saab, Willett, Hochberg, Shenoy, and Henderson]{pandarinath2017high}
Chethan Pandarinath, Paul Nuyujukian, Christine~H Blabe, Brittany~L Sorice, Jad Saab, Francis~R Willett, Leigh~R Hochberg, Krishna~V Shenoy, and Jaimie~M Henderson.
\newblock High performance communication by people with paralysis using an intracortical brain-computer interface.
\newblock \emph{elife}, 6:\penalty0 e18554, 2017.

\bibitem[Pei et~al.(2011)Pei, Barbour, Leuthardt, and Schalk]{pei2011decoding}
Xiaomei Pei, Dennis~L Barbour, Eric~C Leuthardt, and Gerwin Schalk.
\newblock Decoding vowels and consonants in spoken and imagined words using electrocorticographic signals in humans.
\newblock \emph{Journal of neural engineering}, 8\penalty0 (4):\penalty0 046028, 2011.

\bibitem[Poeppel et~al.(2008)Poeppel, Idsardi, and Van~Wassenhove]{poeppel2008speech}
David Poeppel, William~J Idsardi, and Virginie Van~Wassenhove.
\newblock Speech perception at the interface of neurobiology and linguistics.
\newblock \emph{Philosophical Transactions of the Royal Society B: Biological Sciences}, 363\penalty0 (1493):\penalty0 1071--1086, 2008.

\bibitem[Prabhavalkar et~al.(2023)Prabhavalkar, Hori, Sainath, Schl{\"u}ter, and Watanabe]{prabhavalkar2023end}
Rohit Prabhavalkar, Takaaki Hori, Tara~N Sainath, Ralf Schl{\"u}ter, and Shinji Watanabe.
\newblock End-to-end speech recognition: A survey.
\newblock \emph{IEEE/ACM Transactions on Audio, Speech, and Language Processing}, 2023.

\bibitem[Radford et~al.(2022)Radford, Kim, Xu, Brockman, McLeavey, and Sutskever]{radford2022robust}
Alec Radford, Jong~Wook Kim, Tao Xu, Greg Brockman, Christine McLeavey, and Ilya Sutskever.
\newblock Robust speech recognition via large-scale weak supervision. arxiv (2022).
\newblock \emph{arXiv preprint arXiv:2212.04356}, 2022.

\bibitem[Reece H.~Dunn(2015)]{espeakng}
Alexander~Epaneshnikov Reece H.~Dunn, Valdis~Vitolins.
\newblock espeak ng text-to-speech.
\newblock \emph{GitHub}, 2015.
\newblock URL \url{https://github.com/espeak-ng/espeak-ng}.

\bibitem[Schneider et~al.(2019)Schneider, Baevski, Collobert, and Auli]{schneider2019wav2vec}
Steffen Schneider, Alexei Baevski, Ronan Collobert, and Michael Auli.
\newblock wav2vec: Unsupervised pre-training for speech recognition.
\newblock \emph{arXiv preprint arXiv:1904.05862}, 2019.

\bibitem[Schultz \& Wand(2010)Schultz and Wand]{schultz2010modeling}
Tanja Schultz and Michael Wand.
\newblock Modeling coarticulation in emg-based continuous speech recognition.
\newblock \emph{Speech Communication}, 52\penalty0 (4):\penalty0 341--353, 2010.

\bibitem[Touvron et~al.(2023)Touvron, Lavril, Izacard, Martinet, Lachaux, Lacroix, Rozière, Goyal, Hambro, Azhar, Rodriguez, Joulin, Grave, and Lample]{touvron2023llama}
Hugo Touvron, Thibaut Lavril, Gautier Izacard, Xavier Martinet, Marie-Anne Lachaux, Timothée Lacroix, Baptiste Rozière, Naman Goyal, Eric Hambro, Faisal Azhar, Aurelien Rodriguez, Armand Joulin, Edouard Grave, and Guillaume Lample.
\newblock Llama: Open and efficient foundation language models.
\newblock \emph{arXiv preprint arXiv:2302.13971}, 2023.

\bibitem[van~der Maaten \& Hinton(2008)van~der Maaten and Hinton]{JMLR:v9:vandermaaten08a}
Laurens van~der Maaten and Geoffrey Hinton.
\newblock Visualizing data using t-sne.
\newblock \emph{Journal of Machine Learning Research}, 9\penalty0 (86):\penalty0 2579--2605, 2008.
\newblock URL \url{http://jmlr.org/papers/v9/vandermaaten08a.html}.

\bibitem[Vansteensel et~al.(2016)Vansteensel, Pels, Bleichner, Branco, Denison, Freudenburg, Gosselaar, Leinders, Ottens, Van Den~Boom, Van~Rijen, Aarnoutse, and Ramsey]{vansteensel2016fully}
Mariska~J Vansteensel, Elmar~GM Pels, Martin~G Bleichner, Mariana~P Branco, Timothy Denison, Zachary~V Freudenburg, Peter Gosselaar, Sacha Leinders, Thomas~H Ottens, Max~A Van Den~Boom, Peter~C Van~Rijen, Erik~J Aarnoutse, and Nick~F Ramsey.
\newblock Fully implanted brain--computer interface in a locked-in patient with als.
\newblock \emph{New England Journal of Medicine}, 375\penalty0 (21):\penalty0 2060--2066, 2016.

\bibitem[Wei et~al.(2022)Wei, Wang, Schuurmans, Bosma, Xia, Chi, Le, and Zhou]{wei2022chain}
Jason Wei, Xuezhi Wang, Dale Schuurmans, Maarten Bosma, Fei Xia, Ed~Chi, Quoc~V Le, and Denny Zhou.
\newblock Chain-of-thought prompting elicits reasoning in large language models.
\newblock \emph{Advances in neural information processing systems}, 35:\penalty0 24824--24837, 2022.

\bibitem[Willett et~al.(2021)Willett, Avansino, Hochberg, Henderson, and Shenoy]{willett2021high}
Francis~R Willett, Donald~T Avansino, Leigh~R Hochberg, Jaimie~M Henderson, and Krishna~V Shenoy.
\newblock High-performance brain-to-text communication via handwriting.
\newblock \emph{Nature}, 593\penalty0 (7858):\penalty0 249--254, 2021.

\bibitem[Willett et~al.(2023{\natexlab{a}})Willett, Kunz, Fan, Avansino, Wilson, Choi, Kamdar, Glasser, Hochberg, Druckmann, Shenoy, and Henderson]{willett2023high}
Francis~R Willett, Erin~M Kunz, Chaofei Fan, Donald~T Avansino, Guy~H Wilson, Eun~Young Choi, Foram Kamdar, Matthew~F Glasser, Leigh~R Hochberg, Shaul Druckmann, Krishna Shenoy, and Jaimie~M. Henderson.
\newblock A high-performance speech neuroprosthesis.
\newblock \emph{Nature}, 620\penalty0 (7976):\penalty0 1031--1036, 2023{\natexlab{a}}.

\bibitem[Willett et~al.(2023{\natexlab{b}})Willett, Kunz, Fan, Avansino, Wilson, Choi, Kamdar, Glasser, Hochberg, Druckmann, Shenoy, and Henderson]{willett2023highdataset}
Francis~R Willett, Erin~M Kunz, Chaofei Fan, Donald~T Avansino, Guy~H Wilson, Eun~Young Choi, Foram Kamdar, Matthew~F Glasser, Leigh~R Hochberg, Shaul Druckmann, Krishna Shenoy, and Jaimie~M. Henderson.
\newblock Data for: A high-performance speech neuroprosthesis [dataset].
\newblock \emph{Dryad}, pp.\  1--6, 2023{\natexlab{b}}.

\bibitem[Yang et~al.(2024)Yang, Duan, Zhang, Xu, and Xiong]{yang2024decode}
Yiqian Yang, Yiqun Duan, Qiang Zhang, Renjing Xu, and Hui Xiong.
\newblock Decode neural signal as speech.
\newblock \emph{arXiv preprint arXiv:2403.01748}, 2024.

\bibitem[Zhang et~al.(2020)Zhang, Lu, Sak, Tripathi, McDermott, Koo, and Kumar]{zhang2020transformer}
Qian Zhang, Han Lu, Hasim Sak, Anshuman Tripathi, Erik McDermott, Stephen Koo, and Shankar Kumar.
\newblock Transformer transducer: A streamable speech recognition model with transformer encoders and rnn-t loss.
\newblock In \emph{ICASSP 2020-2020 IEEE International Conference on Acoustics, Speech and Signal Processing (ICASSP)}, pp.\  7829--7833. IEEE, 2020.

\bibitem[Zhang et~al.(2022)Zhang, Roller, Goyal, Artetxe, Chen, Chen, Dewan, Diab, Li, Lin, Mihaylov, Ott, Shleifer, Shuster, Simig, Koura, Sridhar, Wang, and Zettlemoyer]{zhang2022opt}
Susan Zhang, Stephen Roller, Naman Goyal, Mikel Artetxe, Moya Chen, Shuohui Chen, Christopher Dewan, Mona Diab, Xian Li, Xi~Victoria Lin, Todor Mihaylov, Myle Ott, Sam Shleifer, Kurt Shuster, Daniel Simig, Punit~Singh Koura, Anjali Sridhar, Tianlu Wang, and Luke Zettlemoyer.
\newblock Opt: Open pre-trained transformer language models.
\newblock \emph{arXiv preprint arXiv:2205.01068}, 2022.

\end{thebibliography}
\bibliographystyle{iclr2025_conference}
%\bibliographystyle{plain}
% \printbibliography
% %\bibliography{iclr2025_conference}

\appendix
\newpage
\section{Appendix}

\subsection{Sensitivity Analysis}
We report the mean and standard deviation of DCoND-L, DCoND-LI and DCoND-LIFT in Table \ref{table:statistical_significance}. The mean and standard deviation are obtained across 5 random seeds. These results show that proposed methods (DCoND-L, DCoND-LI and DCoND-LIFT) performance is robut and they maintain a significant gap over the NPTL and LISA baselines \cite{willett2023high, benster2024cross}.
\begin{table}
  \caption{Sensitivity analysis on Brain-to-Text 2024 Benchmark}
    \centering
  \label{table:statistical_significance}
\begin{tabular}{lllll}
\toprule
 %& \cmidrule(r){1-3} \cmidrule(r){1-3}
          & PER$\times 100\downarrow$ & WER$\times 100\downarrow$ & P-WER$\times 100 \downarrow$ \\
\midrule
NPTL [46] &    16.62  &  9.46   &  11.33                \\       
LISA [2]& ~~-- &  8.93  & ~~-- \\
\midrule
DCoND-L (Ours) &   \textbf{15.44 $\pm$ 0.46}      &     8.39 $\pm$	0.22 &            \textbf{8.09 $\pm$	1.62}         \\
%Tri-phone (5gram + OPT6B) &    15.11      &    9.81     &   31.35    &          6.38        \\
DCoND-LI (Ours) &    ~~--    &     7.23 $\pm$	0.08       &   ~~--        \\
DCoND-LIFT (Ours) &    ~~--      &    \textbf{5.90 $\pm$	0.08}          &  ~~--          \\
\bottomrule
\end{tabular}
\end{table}

\subsection{Triphone as an alternative for context-dependent phoneme representation}\label{appendex:triphone}
 % \begin{itemize}
 %     \item possible ways to define triphone.
 %     \item number of classes.
 % \end{itemize}
Triphones expand upon diphones by incorporating a larger context. Specifically, a triphone considers one phoneme before and one phoneme after the current main phoneme. Consequently, when a neural signal segment is decoded into acoustic units based on the continuity of three phonemes, it reflects a triphone structure.  For example, the single phoneme sequence 
\[H,~~~OW,~~~P \] for ``hope", can be transferred to triphone \[``SIL\rightarrow H \rightarrow OW,~~~H\rightarrow OW \rightarrow P,~~~OW \rightarrow P\rightarrow SIL".\]
In this scenario, the time steps required for decoding single phonemes and triphones remain the same. However, triphones introduce a substantial increase in the number of classes, \textbf{scaling as $N^3$}, which can be prohibitively large (e.g., $64000$ when $N=40$). The divide and conquer approach in this case is expressed as
\[
f(x) = p(Z=c_i|X) = \sum_{c_j\in C, c_q \in C} p(c_j, c_i, c_q | X)
\]
Similarly to the diphone probability matrix, the triphone classes are then mapped into a three dimensional triphone matrix (tensor), where each element represents the probability of the current neural signal encoding the phoneme transition from phoneme $c_j$ to phoneme $c_i$ and concluding at phoneme $c_q$. By summing over the first and last dimensions, we obtain $p(Z=c_i|X)$. Given the potential sparsity of triphone combinations, certain triphone subclasses may not occur frequently in a given language. To mitigate this, we select the top $K$ subclasses for each triphone sample based on occurrence counts within the current vocabulary. Specifically, for a main phoneme $c_i$, we rank all possible combinations of $*->c_i->*$ and retain the top $K$ as subclasses for the phoneme class $c_i$.

Aside from selecting the top $K$ subclasses, an alternative approach involves grouping phones according to articulation similarity \cite{herff2015brain}. This categorization leads to subclasses of the phoneme $c_i$ as $group_j->c_i->group_q$. We categorize phonemes into 14 groups, encompassing Bilabial Sounds, Labiodental Sounds, Dental Sounds, Alveolar Sounds, Palatal Sounds, Velar Sounds, Glottal Sounds, Front Vowels, Central Vowels, Back Vowels, and SIL. In this context, the number of subclasses amounts to $14*40*14$, which is comparable to the number of classes when $K=200$ (resulting in a total of 200*40 subclasses).
\subsection{Additional Ablation Study on the Contribution of LMs}\label{appendix:ablation_llm}
\begin{figure*}
\begin{center}
 \includegraphics[width=0.5\linewidth]{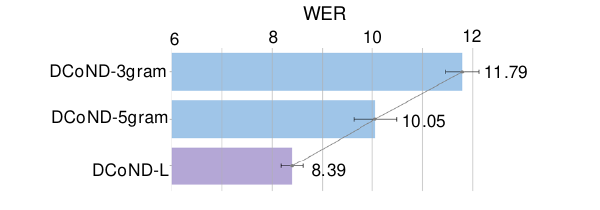}
\end{center}
 \vspace{-5pt}
   \caption{Ablation study on the contribution of re-scoring step in the phoneme-to-transcription pipeline.
 }
\label{fig:Fig5-wer-ablation-appendix}
\end{figure*}

We conduct an additional study to assess the role of phoneme-to-transcription generation and re-scoring methods (Figure \ref{fig:Fig5-wer-ablation-appendix}). We show that removing the re-scoring step performed by the OPT model in DCoND-L significantly degrades WER (DCoND-3gram and DCoND-5gram), highlighting the importance of the transcription re-scoring step. In addition, the 5-gram model with longer phoneme dependency generates more accurate transcription candidates compared to the 3-gram model.
%\begin{figure}{r}{1\textwidth}
%\end{wrapfigure}

\subsection{Open-Source LLMs for DCoND-LI \& DCoND-LIFT}
In addition to the closed-source GPT-3.5, we explored the use of the open-source Llama-3.1-70B for refining transcription predictions. We evaluated Llama-3.1-70B in both in-context learning (DCoND-LI) and fine-tuning (DCoND-LIFT) scenarios and compare it against GPT3.5 (Table \ref{table:opensource-llm}). Llama-3.1-70B performs on par with GPT3.5 in ICL setting, while closely trail behind in the finetuning setting. Either variants appear to outperform NPTL and LISA baselines. These results demonstrate DCoND method’s robustness and generalizability to other LLMs besides GPT3.5, and warrant the accessibility of our methods to the broad community.

\begin{table}
\centering
\caption{GPT-3.5 vs Llama-3.1-70B for error correction from ensemble of transcriptions}
\label{table:opensource-llm}
\begin{tabular}{lll}
\toprule
          &  Llama-3.1-70B WER          & GPT 3.5 WER                  \\
        \midrule
 DCoND-LI  & 7.38                        &  7.29                   \\
DCoND-LIFT & 6.85 & 5.77\\
\bottomrule
\end{tabular}
\end{table}
\subsection{Investigation on Architecture Choices for Neural Decoders}\label{appendix:model-architecture}
We study the effects of different model architectures on the phoneme decoding performance (PER) (Table \ref{table:model-architecture}). We observe a significant performance degradation in PER when using Transformer as the neural decoder. On the other hand, RNN counterparts (LSTM and GRU) perform decently well, with GRU being the most performant model for both single phoneme decoding (NPTL) and diphone decoding (DCoND).
\begin{table}
\centering
\caption{Comparison of different model architectures on phoneme decoding performance} \label{table:model-architecture}
\begin{tabular}{llll}
\toprule
      & \multicolumn{3}{c}{PER}    \\
      \cmidrule(l){2-4}
      & Transformer               & LSTM &\textbf{GRU} \\
NPTL  &  39.58 $\pm$ 0.15 & 17.49 $\pm$ 0.32 & \textbf{16.63 $\pm$ 0.19}  \\
DCoND & 38.88 $\pm$ 0.17 &16.08 $\pm$ 0.23 & \textbf{15.44 $\pm$ 0.46} \\              \bottomrule
\end{tabular}
\end{table}
\subsection{Implementation Details}\label{appendix:implementation detail}
We preprocess the recorded neural signals and construct an RNN neural encoder following the methodology outlined in \cite{willett2023high}. The raw neural signals $X \in \mathbb{R}^{T\times D}$ are initially partitioned into smaller patches with a window size of $W$, where each patch of shape $X \in \mathbb{R}^{T'\times (DW)}$ can overlap and the overlap between patches is determined by the stride size $W$. Stride size is set to $W=14$ for diphone experiments and $W=32$ for the triphone experiemnts.
The bidirectional RNN processes these patched neural signals as inputs, which are subsequently transformed into the neural representation space $H=[h_1, h_2, \cdots, h_{T'}] \in \mathbb{R}^{T' \times d }$. A fully connected layer then maps hidden representations to diphone or triphone subclasses, denoted as $P(S=s_i|X)$. The outputs of the fully connected layer are used to compute $\mathcal{L}_s$. 
The computation of single phoneme probabilities is detailed in Equation \ref{sec:diphone collapse}. We merge the probability computed from diphone or triphone. 

During the RNN training, we utilize a batch size of 32, a learning rate of 0.02, and the Adam optimizer across various experiments the same set of parameters as used in NPTL baseline \cite{willett2023high}. To facilitate diphone and triphone learning, we initially train the subclasses for 10 epochs and then gradually increase the ratio of the single phoneme loss by 0.1 every 10 epochs until it reaches 0.6.
The number of training epochs varies across single phoneme learning, diphone learning, and triphone learning. Specifically, we conduct experiments for up to 100 epochs for single phoneme learning (NPTL baseline), 120 epochs for diphone learning, and 140 epochs for triphone learning since the diphone and triphone include subclasses and require additional subclass training procedures. Increasing the number of training epochs can often lead to overfiting of the training data. Training was done on 2 GeForce RTX 2080 Ti with around 12GB memory. The training typically took around 6-8 hours.

The 5-gram model takes the predicted phoneme logits as inputs, which can be scaled by a temperature factor denoted as $t$ using the formula $logits := logits/t$. Through experimentation, we have found that setting $t=1.2$ improves the decoding performance. Therefore, we use $t=1.2$ for our experiments, including the implementation of NPTL, which has resulted in improved baseline results. Specifically, the leaderboard score has improved from 9.76 to 9.46.

% We preprocess the neural signal and build RNN neural encoder as described in the paper \cite{willett2023high}. Raw neural signal $X \in R^{T\times D}$ is first patched into small patches with window size $w$. The patched neural signal has shape $X \in R^{T'\times (DW)}$. Overlapping between the patches is allowed and depends on stride size. 
% The bidirectional RNN takes the patched neural signal as inputs which are then transformed into the neural representation space $H=[h_1, h_2, \cdots, h_{T'}] \in R^{T' \times d }$. A fully connected layer maps the hidden representation to diphone or tiphone subclasses $P(S=s_i|X)$. From the distribution of diphone or triphone the single phoneme probability is computed as described in \ref{sec:diphone collapse}. During the training of RNN we use batch size 32, learning rate 0.02, and Adam optimizer across different experiments. The number of training epochs could be different for single phoneme learning, diphone learning and triphone learning. We run the experiment up to 100 epochs during single phoneme learning, 120 during the diphone learning and 140 during the triphone learning. To learn the diphone and triphone, we first learn the subclasses for 10 epochs, then we gradually increase the ratio of the single phoneme loss over until it reaches 0.6.
\subsection{Phoneme Error Analysis}
%\begin{wrapfigure}{r}{0.5\textwidth}
\begin{figure}
\centering
\includegraphics[width=0.5\linewidth]{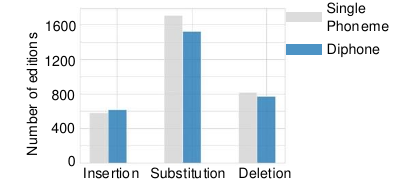}
   \caption{Phoneme error types analysis during single phoneme decoding and diphone.
 }
\label{fig:Fig6-appendix-phonemeerrorbar}
\end{figure}
%\end{wrapfigure}
We conducted a detailed analysis of the various types of errors encountered during phoneme decoding. This analysis involved assessing the operations necessary to align the decoded phoneme sequence with the ground truth phonemes, comparing scenarios where only single phoneme decoding is used versus employing diphone subclass decoding.
Overall, our findings indicated that employing diphone subclass decoding leads to a reduction in the number of operations required to align the decoded sequence with the ground truth phonemes. Specifically, fewer editing operations, particularly substitutions, are needed when utilizing the diphone decoding paradigm compared to directly decoding single phonemes.

\subsection{Prompt for GPT3.5}\label{appendix:gptprompt}
\paragraph{Prompt used for to GPT3.5 is as follows:} Your task is to perform automatic speech recognition. Below are multiple candidate transcriptions together with their corresponding phoneme representations. The phonemes are taken from the CMU Pronouncing Dictionary. The special symbol SIL represents the start of the sentence, or the end of the sentence, or the space between two adjacent words. Based on the transcription candidates and their phoneme representations, come up with a transcription and its corresponding phoneme representation that are most accurate, ensuring the transcription is contextually and grammatically correct. Focus on key differences in the candidates that change the meaning or correctness. Avoid selections with repetitive or nonsensical phrases. In cases of ambiguity, select the option that is most coherent and contextually sound, taking clues from the phoneme representations. The candidate phoneme representations may not always be the correct representation of the corresponding candidate transcriptions. Some phonemes in the candidate phoneme sequences might have been incorrectly added, removed, or replaced. However, the candidate phonemes contain useful information that will help you come up with the correct transcription and phoneme representation. You should translate each subgroup of phonemes that is enclosed by two SIL symbols into one single word. You should remove SIL symbols at the start or the end of the phoneme sequence. Respond with your refined transcription and its corresponding phoneme representation only, without any introductory text.

\paragraph{Examples of prediction and correction pairs}
Examples of prediction and correction pairs are presented in Table \ref{table1:example-of-ft}. Additionally, we show In-Context-Learning prompts in Table \ref{table1:example-of-icl}.
\begin{table}
\centering
  \caption{Examples of prediction and ground truth (GT) pairs.}
    \centering
  \label{table1:example-of-ft}
\begin{tabularx}{\textwidth}{|lX|}
\toprule
Transcription candidate 1: &but we don't know that.\\ Transcription candidate 2: &but we don't know that.\\ Transcription candidate 3: &but you don't know that.\\Transcription candidate 4: &but you don't know that.\\ Transcription candidate 5: &but you don't know that.\\ Transcription candidate 6: &but you don't know that.\\ Transcription candidate 7: &but you don't know that.\\ Transcription candidate 8: &but you don't know that.\\ Transcription candidate 9: &but we don't know that.\\Transcription candidate 10: &but we don't know that.\\ 
\midrule
Phoneme candidate 1: &SIL B AH T SIL W IY SIL D OW N T SIL N OW SIL DH AE T SIL.\\ 
Phoneme candidate 2:& SIL B AH T SIL Y IY SIL D OW N T SIL N OW SIL DH AE T SIL.\\ 
Phoneme candidate 3: &SIL B AH T SIL Y UW SIL D OW N T SIL N OW SIL AE T SIL.\\ 
Phoneme candidate 4: &SIL B AH T SIL Y UW SIL D OW N T SIL N OW SIL DH AE T SIL.\\ 
Phoneme candidate 5: &SIL B AH T SIL DH UW SIL D OW N T SIL N OW SIL DH AE T SIL.\\ 
Phoneme candidate 6: &SIL B AH T SIL Y UW SIL D OW N T SIL N OW SIL DH AE T SIL.\\ 
Phoneme candidate 7: &SIL B AH T SIL Y UW SIL D OW N T SIL N OW SIL DH AE T SIL.\\ 
Phoneme candidate 8: &SIL B AH T SIL Y UW SIL D OW N T SIL N OW SIL DH AE T SIL.\\ 
Phoneme candidate 9: &SIL B AH T SIL W IY SIL D OW N T SIL N OW Z SIL DH AE T SIL.\\ 
Phoneme candidate 10: &SIL B AH T SIL DH IY SIL D OW N T SIL N OW SIL AE T SIL. \\
\midrule
\textbf{Transcription GT}:  &but we don't know that.\\
\textbf{Phoneme GT}: &SIL B AH T SIL W IY SIL D OW N T SIL N OW SIL DH AE T SIL. \\
 \bottomrule
\end{tabularx}
\end{table}

\begin{table}
\centering
  \caption{Example of In-Context-Learning (ICL) prompts and query.}
    \centering
  \label{table1:example-of-icl}
\begin{tabularx}{\textwidth}{|lX|}
\toprule
\textcolor{red}{System Prompt: } & Your task is to perform automatic speech recognition. You are given ten candidates of an unknown transcription. Your job is to come up with a transcription that is most accurate, relying on the context that the candidates provide. First, observe the provided examples demonstrating how the task should be done, then work on the query candidates. In each example, ten transcription candidates, their corresponding phoeneme representations, and a ground truth transcription are given. The ground truth transcription is the correct transcription, while the transcription candidates and phoneme representations may or may not contain errors. Some phonemes in the phoneme sequences might have been incorrectedly added, removed, or replaced. However, the phonemes contain helpful information that will help you come up with the correct transcription. You should translate each subgroup of phonemes that is enclosed by two SIL symbols into one single word. You should remove SIL symbols at the start and the end of the phoneme sequence. Make sure your transcription based on the query candidates is contextually and grammatically correct. Focus on key differences in the candidates that change the meaning or correctness. Avoid selections with repetitive or nonsensical phrases. In cases of ambiguity, select the option that is most coherent and contextually sound. Respond with your final transcription only, without any introductory text. \\
\textcolor{blue}{Context prompt}:    & \textbf{Example 1}: \textit{Transcription candidate 1}: i enjoyed it very much. $\cdots$ \textit{Transcription candidate 10:} i enjoyed it very much. \textit{Phoneme candidate 1:} AY SIL EH N JH OY D SIL IH T SIL V EH R IY SIL M AH CH SIL. $\cdots$ \textit{Phoneme candidate 10:} AY SIL EH N JH OY D SIL IH T SIL V EH R IY SIL M AH CH SIL. $\cdots$ \textbf{Ground truth phonemes:} AY SIL EH N JH OY D SIL IH T SIL V EH R IY SIL M AH CH. \textbf{Ground truth transcription}: i enjoyed it very much. $\cdots$ 
\textbf{Example N:} \textit{Transcription candidate 1: }the ranks of asian riders are falling too. $\cdots$ \textit{Transcription candidate candidate 10: }the ranks of asian riders are willing to. \textit{Phoneme candidate 1}: DH AH SIL R AE NG K S SIL AH V SIL EY ZH AH N SIL R AY D Z SIL AA R SIL F L D IH NG SIL T UW SIL. $\cdots$ \textit{Phoneme candidate 10}: DH AH SIL R AE K S SIL AH V SIL EY ZH AH N SIL R EY D ER Z SIL AA R SIL F IY L IH NG SIL T UW SIL. \textbf{Ground truth phonemes}: DH AH SIL R AE NG K S SIL AH V SIL EY ZH AH N SIL R AY D ER Z SIL AA R SIL S W EH L IH NG SIL T UW. \textbf{Ground truth transcription}: the ranks of asian riders are swelling too \\
\midrule

 \textcolor{green}{Query}: & Transcription candidate 1: i'm originally from colorado. $\cdots$ Transcription candidate 10: i'm only from colorado. Phoneme candidate 1: SIL AY M SIL ER N AH L IY SIL F R AH M SIL K AO L ER AA D OW SIL. $\cdots$ Phoneme candidate 10: SIL AY M SIL AH N L IY SIL F R AH M SIL K AO L R AA D OW SIL. \\
 \bottomrule
\end{tabularx}
\end{table}

% \subsection{Effect of Alpha}
% See Table\ref{table:ablation-alpha}
% \begin{table}[]
%     \caption{Ablation study for different definitions of phoneme subclasses.}
%     \label{table:ablation-alpha}
%     \centering
%     \begin{tabular}{lllll}
%     \toprule
%     alpha value 0.0 & 0.2   & 0.4   & 0.6 & 0.8 & 1 \\
%     per         & 15.8  & 15.49 &                         &                         \\
%     wer         & 12.58 & 12.19 &                         &                 \\
%     \bottomrule
%     \end{tabular}
% \end{table}

\end{document}